\documentclass[useAMS,usenatbib]{mn2e}

\usepackage{times}
\usepackage{footmisc}
\usepackage{amsmath}
\usepackage{graphicx}
\usepackage{lscape}
\usepackage{psfig}

\title[MSP timing stability and GW detection]{Timing stability of
  millisecond pulsars and prospects for gravitational-wave detection}

\author[J.~P.~W.~Verbiest, et al.]
  {J.~P.~W.~Verbiest,$^{1,2,3}$\thanks{E-mail: Joris.Verbiest@mail.wvu.edu.}
  M.~Bailes,$^1$ W.~A.~Coles,$^4$ G.~B.~Hobbs,$^2$ W.~van Straten,$^1$
  \newauthor D.~J.~Champion,$^2$ F.~A.~Jenet,$^5$ R.~N.~Manchester,$^2$
  N.~D.~R.~Bhat,$^1$\newauthor J.~M.~Sarkissian,$^2$ D.~Yardley,$^{2,6}$
  S.~Burke-Spolaor,$^{1,2}$ A.~W.~Hotan$^7$ and X.~P.~You$^8$\\
  $^1$Swinburne University of Technology, Centre for Astrophysics and
  Supercomputing Mail \#H39 P. O. Box 218 VIC 3122
  Australia\\
  $^2$Australia Telescope National Facility $-$ CSIRO P.O. Box 76,
  Epping, NSW, 1710
  Australia\\
  $^3$Department of Physics, West Virginia University, P.O. Box 6315,
  WV 26506, USA\\
  $^4$Electrical and Computer Engineering, University of California at
  San Diego, La Jolla, CA 92093, USA\\
  $^5$CGWA, University of Texas at Brownsville, TX 78520, USA\\
  $^6$Sydney Institute for Astronomy, School of Physics A29, The
  University of Sydney, NSW 2006, Australia\\
  $^7$Curtin Institute for Radio Astronomy, Curtin University of
  Technology, Bentley, WA 6102, Australia\\
  $^8$School of Physical Science and Technology, Southwest University,
  2 Tiansheng Road, Chongqing 400715, China}
\date{Submitted 2009 Jun 19}

\begin{document}

\label{firstpage}

\maketitle

\begin{abstract}
Analysis of high-precision timing observations of an array of $\sim$20
millisecond pulsars (a so-called ``timing array'') may ultimately
result in the detection of a stochastic gravitational-wave
background. The feasibility of such a detection and the required
duration of this type of experiment are determined by the achievable
rms of the timing residuals and the timing stability of the pulsars
involved. We present results of the first long-term, high-precision
timing campaign on a large sample of millisecond pulsars used in
gravitational-wave detection projects. We show that the timing
residuals of most pulsars in our sample do not contain significant
low-frequency noise that could limit the use of these pulsars for
decade-long gravitational-wave detection efforts.  For our most
precisely timed pulsars, intrinsic instabilities of the pulsars or the
observing system are shown to contribute to timing irregularities on a
five-year timescale below the 100\,ns level.  Based on those results,
realistic sensitivity curves for planned and ongoing timing array
efforts are determined. We conclude that prospects for detection of a
gravitational-wave background through pulsar timing array efforts
within five years to a decade are good.
\end{abstract}


\section{Introduction}\label{sec:intro}
The rotational behaviour of pulsars has long been known to be
predictable, especially in the case of millisecond pulsars (MSPs).
Current models suggest that such pulsars have been spun up by
accretion from their binary companion star to periods of several
milliseconds, making them spin much faster than the more numerous
younger pulsars, which typically have periods of about a second. The
rotational stability of MSPs is generally 3-4 orders of magnitude
better than that of normal pulsars and on timescales of several years,
it has been shown that some MSPs have a timing stability comparable to
the most precise atomic clocks \citep{mte97}. This timing stability is
most clearly quantified through the technique of pulsar timing, which
compares arrival times of pulses to a model describing the pulsar, its
binary orbit and the interstellar medium (ISM) between the pulsar and
Earth \citep[as detailed by][]{ehm06}. This technique has enabled
determination of physical parameters at outstanding levels of
precision, such as the orbital characteristics of binary star systems
\citep[e.g.][]{vbb+01}, the masses of pulsars and their companions
\citep[e.g.][]{jhb+05,nic06} and the turbulent character of the ISM
\citep[e.g.][]{yhc+07}. The strong gravitational fields of pulsars in
binary systems have also enabled stringent tests of general relativity
(GR) and alternative theories of gravity, as described by, e.g.,
\citet{ksm+06} and \citet{bbv08}. Finally, pulsars have provided the
first evidence that gravitational waves (GWs) exist at levels
predicted by GR \citep{tw82} and have placed the strongest limit yet
on the existence of a background of GWs in the Galaxy \citep{jhv+06}.

\citet{saz78} was the first to investigate the potential effect of GWs
on the times-of-arrival (TOAs) of pulsar pulses and to conclude that
direct detection of GWs could be possible through pulsar
timing. Subsequent analyses and theoretical predictions for
astronomical sources of GWs have determined that a stochastic
gravitational-wave background (GWB) from binary black holes in the
cores of galaxies is the most likely signal to be detectable. As
summarised in \citet{jhv+06}, the energy density of such a GWB per
unit logarithmic frequency interval can be expressed as:
\begin{equation}\label{eq:Omega}
  \Omega_{\rm gw}(f) = \frac{2}{3}\frac{\pi^2}{H_{\rm 0}^2} A^2
	\frac{f^{2\alpha+2}}{f_{\rm ref}^{2\alpha}},
\end{equation}
where $H_{\rm 0} = 100 h$\,km s$^{-1}$ Mpc$^{-1}$ is the Hubble
constant, $f$ is the GW frequency, $f_{\rm ref} = (1\,{\rm yr})^{-1}$,
$A$ is the dimensionless amplitude of the GWB and $\alpha$ is the
spectral index of the GWB. The one-sided power spectrum of the effect
of such a GWB on pulsar timing residuals is given by:
\begin{equation}\label{eq:GWBPower}
  P(f) = \frac{1}{12 \pi^2 f^3} h_{\rm c}(f)^2,
\end{equation}
where $h_{\rm c}$ is the characteristic strain spectrum, defined as:
\begin{equation}\label{eq:h_c}
  h_{\rm c}(f) = A \left(\frac{f}{f_{\rm ref}}\right)^{\alpha}.
\end{equation}

\citet{jhv+06} also summarized the characteristics and expected ranges
for various GWBs of interest. Most importantly, for a GWB created by
supermassive black-hole mergers, $\alpha = -2/3$ and $A$ is predicted
to be between $10^{-15}$ and $10^{-14}$
\citep{jb03,wl03a}. \citet{svc08} expanded upon these analyses and
showed that the actual GWB spectrum strongly depends on the merger
history, with a variety of spectral indices possible. They concluded,
however, that $\alpha = -2/3$ was a reasonable approximation for
practical purposes. For a background of GWs that were formed in the
early Universe, $\alpha \approx -1$ and the amplitude range predicted
by \citet{gri05} is $A = 10^{-17} - 10^{-15}$, but standard models
\citep[e.g.][]{bb08} predict much lower amplitudes. A third GWB that
may be detected by PTAs, is formed by cosmic strings
\citep{dv05,cbs96}, with predicted amplitudes between $10^{-16} -
10^{-14}$ and $\alpha = -7/6$ \citep{mag00}.

\citet{hd83} first investigated the correlations that arise between
timing data of different pulsars due to the presence of a stochastic
and isotropic GWB in the Galaxy. They demonstrated that the GWs cause
a quadrupolar correlation between the timing residuals of different
pulsars. \citet{rom89} and \citet{fb90} expanded this analysis and
introduced the concept of a pulsar timing array (PTA), in which an
ensemble of pulsars is timed and their residuals correlated with each
other. The PTA concept uses the quadrupolar correlation signature
first derived by \citet{hd83} to separate the effect of a GW from all
other contributions to the residuals, such as intrinsic pulsar timing
irregularities, clock errors, ISM effects and Solar System ephemeris
errors. Alternatively, the correlation signature for non-Einsteinian
GWs \citep[as recently derived by][]{ljp08} could be used.

The PTA concept was more rigorously explored by \citet{jhlm05} who
first determined the sensitivity of PTA experiments to backgrounds of
GWs \citep[Equation (12) of][]{jhlm05}. Their analysis showed that the
sensitivity of a PTA depends on four main parameters: the number of
pulsars, the data span ($T$), the root-mean-square of the timing
residuals (simply `rms' or $\sigma$ henceforth) and the number of
observations in each of the pulsar timing data sets ($N_{\rm
  TOA}$). They further determined that, for a PTA consisting of weekly
observations of 20 MSPs, all with a timing rms of 100\,ns, a five-year
observational campaign would be required to make a $\sim3\,\sigma$
detection of a GWB with $\alpha = -2/3$ and $A = 10^{-15}$. It follows
from Equation (12) of \citet{jhlm05} that the lowest amplitude of a
GWB from supermassive black-hole mergers to which a PTA is sensitive,
scales as:
\begin{equation}\label{eq:ScalingLaw}
  A_{\rm min, GWB} \propto \frac{\sigma}{T^{5/3}\sqrt{N_{\rm TOA}}}
\end{equation}
Depending on the achievable rms residual of MSPs, an alternative PTA
could therefore achieve the same results through timing of 20 MSPs on
a biweekly basis for ten years with an rms of close to 300\,ns. This
raises two questions related to the potential of PTAs to detect a
GWB. First, down to which precision can MSPs be timed ($\sigma_{\rm
  min}$) and second, can a low residual rms be maintained over long
campaigns (i.e. does $\sigma/T^{5/3}$ decrease with time)? In the
context of this second question, we will henceforth use the term
``timing stability'' when referring to the potential of an MSP timing
data set to maintain a constant, preferably low rms residual at all
timescales up to the timespan of a PTA project, which is typically
envisaged to be five years or longer.

It has been shown for a few pulsars that timing with a residual rms of
a few hundred nanoseconds is possible for campaigns lasting a few
years. Specifically, \citet{hbo06} presented a timing rms of 200\,ns
over two years of timing on PSRs J1713+0747 and J1939+2134 (PSR
B1937+21) and 300\,ns over two years of timing on PSR J1909$-$3744;
\citet{sns+05} reported an rms of 180\,ns on six years of timing PSR
J1713+0747 and \citet{vbv+08} timed PSR J0437$-$4715 at 200\,ns over
ten years. Similar results for PSRs J0437$-$4715 and J1939+2134 were
obtained by \citet{hbb+08} over five years of timing. It has, however,
not been demonstrated thus far that MSPs can be timed with an rms
residual of $\le 100$\,ns over five years or more.

The second question - whether a low rms residual can be maintained
over ten years or longer, also remains unanswered. \citet{ktr94}
detected excess low-frequency noise in PSR J1939+2134; \citet{sns+05}
presented apparent instabilities in long-term timing of PSR J1713+0747
and \citet{vbv+08} noted correlations in the timing residuals of PSR
J0437$-$4715, but apart from these, no long-term timing of MSPs has
been presented to date. Given the low rms residual reported on all
three sources, it is unclear how strongly the reported non-Gaussian
noise would affect the use of these pulsars in a GWB detection effort.

In this article we present the first high-precision long-term timing
results for a sample of 20 MSPs. The source selection, observing
systems and data analysis methods are described in
\S\ref{sec:obs}. Our updated timing models and residual plots for all
pulsars in our sample are also presented in \S\ref{sec:obs}, allowing
the reader a fundamental inspection of the reliability of our timing
results. In \S\ref{sec:stab}, we perform a stability analysis of the
timing data, with the dual purpose of identifying low-frequency noise
in any of our timing data and of assessing the potential impact of
such noise on the use of pulsars in a timing array. In
\S\ref{sec:stability} we outline a new way of quantifying different
components of the pulsar timing rms. Through this analysis, we
separate the levels of receiver noise, noise with a dependency on
observing frequency and any temporal instabilities, providing a bound
on the residual rms that might be achievable on a five-year
timescale. We apply this analysis to three of our most precisely timed
pulsars. In \S\ref{sec:PTA}, we calculate sensitivity curves for
ongoing and planned PTAs. These sensitvity curves take into account
the inhomogeneous character of a realistic array (i.e. the fact that
the rms will differ between pulsars) and assume a bound on residual
rms as determined in \S\ref{sec:stability}. In \S\ref{sec:conclusions}
we summarise our findings.

\section{Observations and Data Reduction}\label{sec:obs}
\subsection{Sample Selection}
The data presented in this paper have been collated from two pulsar
timing programmes at the Parkes radio telescope. The oldest of these
commenced during the Parkes 70\,cm MSP survey \citep{bhl+94}, aiming
to characterise properly the astrometric and binary parameters of the
MSPs found in the survey. Initial timing results from this campaign
were published by \citet{bbm+97} and \citet{tsb+99}. The bright
millisecond pulsars PSRs J1713+0747 and B1937+21 (both discovered
earlier at Arecibo) were also included in this programme. A few years
later, as new discoveries were made in the Swinburne
intermediate-latitude survey \citep{ebvb01}, these pulsars were also
added, resulting in a total of 16 MSPs that were regularly timed by
2006. Improved timing solutions for these 16 pulsars were presented by
\citet{hbo06} and \citet{ojhb06}.

Besides the projects described above, the Parkes Pulsar Timing Array
\citep[PPTA;][]{man08} project commenced more regular timing
observations of these pulsars in late 2004, expanding the number of
MSPs to 20 (listed in Table \ref{tab:psrs}) and adding regular
monitoring at a low observing frequency ($685$\,MHz) to allow
correction for variations of the ISM electron density. A detailed
analysis of these low frequency observations and ISM effects was
recently presented by \citet{yhc+07} and an analysis of the combined
data on PSR J0437$-$4715 was published by \citet{vbv+08}. For this
pulsar we will use the timing results presented in that publication;
for all other pulsars we will present our improved timing models in
\S\ref{sec:timing}.

\begin{table*}
  \begin{center}
	\begin{minipage}{17.5cm}
  \caption{Pulsars in our sample. Column 2 gives the reference for the
  discovery paper, while column 3 provides references to recent or
  important publications on timing of the sources. For the three
  pulsars with original B1950 names, these names are given beside the
  J2000.0 names.}
  \label{tab:psrs}
	\begin{tabular}{l|l|r|r|r|r}
	  \hline
	  \multicolumn{1}{c}{Pulsar} & \multicolumn{1}{c}{Discovery} & 
    \multicolumn{1}{c}{Previous} & \multicolumn{1}{c}{Pulse} & 
    \multicolumn{1}{c}{Orbital} & \multicolumn{1}{c}{Dispersion} \\ 
	  \multicolumn{1}{c}{name} &  &  \multicolumn{1}{c}{timing}
	  & \multicolumn{1}{c}{period (ms)} & \multicolumn{1}{c}{period (d)}
    & \multicolumn{1}{c}{measure}\\
	  & & \multicolumn{1}{c}{solution\footnote{References: (1)
	  \citet{vbv+08}; (2) 
	  \citet{vbb+01}; (3) \citet{hbo06}; (4) \citet{tsb+99}; (5)
	  \citet{ojhb06}; (6) \citet{sns+05}; (7) \citet{eb01b}; (8)
	  \citet{hlk+04}; (9) \citet{ktr94}; (10) \citet{cb04}; (11)
	  \citet{jhb+05}; (12) \citet{lkd+04}}} & & &
	  \multicolumn{1}{l}{(cm$^{-3}$ pc)} \\ 
	  \hline
	  J0437--4715 & \cite{jlh+93} & 1, 2 &  5.8 &   5.7 &  2.6\\
	  J0613--0200 & \cite{lnl+95} & 3  &  3.1 &   1.2 & 38.8\\
	  J0711--6830 & \cite{bjb+97} & 3, 4  &  5.5 &   --  & 18.4\\ 
	  J1022+1001  & \cite{cnst96} & 3  & 16.5 &   7.8 & 10.3\\
	  J1024--0719 & \cite{bjb+97} & 3  &  5.2 &   --  &  6.5\\
	  \\
	  J1045--4509 & \cite{bhl+94} & 3  &  7.5 &   4.1 & 58.2\\
	  J1600--3053 & \cite{ojhb06} & 5  &  3.6 &  14.3 & 52.3\\
	  J1603--7202 & \cite{llb+96} & 3  & 14.8 &   6.3 & 38.0\\
	  J1643--1224 & \cite{lnl+95} & 4  &  4.6 & 147.0 & 62.4\\
	  J1713+0747  & \cite{fwc93}  & 3, 6  &  4.6 &  67.8 & 16.0\\
	  \\
	  J1730--2304 & \cite{lnl+95} & 4 &  8.1 &   --  &  9.6\\
 	  J1732--5049 & \cite{eb01b}  & 7  &  5.3 &   5.3 & 56.8\\
	  J1744--1134 & \cite{bjb+97} & 3  &  4.1 &   --  &  3.1\\
	  B1821--24; J1824$-$2452 & \cite{lbm+87} & 8, 10 &  3.1 &   --  &120.5\\
	  B1855+09; J1857+0943 & \cite{srs+86} & 3, 9  &  5.4 &  12.3 & 13.3\\
	  \\
	  J1909--3744 & \cite{jbv+03} & 3, 11  &  2.9 &   1.5 & 10.4\\
	  B1937+21; J1939+2134 & \cite{bkh+82} & 3, 9  &  1.6 &   --  & 71.0\\
	  J2124--3358 & \cite{bjb+97} & 3  &  4.9 &   --  &  4.6\\
	  J2129--5721 & \cite{llb+96} & 3  &  3.7 &   6.6 & 31.9\\
	  J2145--0750 & \cite{bhl+94} & 3, 12  & 16.1 &   6.8 &  9.0\\
	  \hline
	\end{tabular}
	\end{minipage}
  \end{center}
\end{table*}

\subsection{Observing Systems}\label{sec:obsSys}
Unless otherwise stated, the data presented were obtained at the
Parkes 64\,m radio telescope, at a wavelength of 20\,cm. Two receivers
were used: the H-OH receiver and the 20\,cm multibeam receiver
\citep{swb+96}. Over the last five years, observations at 685\,MHz
were taken with the 10/50\,cm coaxial receiver for all pulsars;
however, they were only used directly in the final timing analysis of
PSR J0613$-$0200, whose profile displays a sharp spike at this
frequency if coherent dedispersion is applied. For PSRs J1045$-$4509,
J1909$-$3744 and J1939+2134, the 685\,MHz observations were used to
model and remove the effects of temporal variations in interstellar
dispersion delays and hence are included indirectly in the timing
analysis. For all other pulsars any such variations were below the
level of our sensitivity.

Three different observing observing systems systems were
used. Firstly, the Caltech Fast Pulsar Timing Machine
\citep[FPTM;][]{sbm+97,san01}, between 1994 and November 2001. This is
an autocorrelation spectrometer with a total bandwidth of up to
256\,MHz. Secondly, the 256\,MHz bandwidth analogue filterbank (FB)
was used in 2002 and 2003. Finally, the Caltech-Parkes-Swinburne
Recorder 2 \citep[CPSR2;][]{hbo06} was used from November 2002
onwards. CPSR2 is a baseband data recorder with two 64\,MHz bandwidth
observing bands (one usually centred at 1341\,MHz, the other at
1405\,MHz) and phase-coherent dispersion removal occuring in near real
time.

\subsection{Arrival Time Determination}\label{ssec:ATD}
The processing applied differs for data from different observing
systems. The FPTM data were calibrated using a real-time system to
produce either two or four Stokes parameters which were later combined
into Stokes I. The FB data were produced from a search system with no
polarimetric calibration possible. This system produced Stokes I
profiles after folding 1-bit data. Data from both of these systems
were integrated in frequency and time to produce a single profile for
each observation. These observations were $\sim$25\,minutes in
duration. For CPSR2 data, in order to minimise the effects of aliasing
and spectral leakage, 12.5\% of each edge of the bandpass was
removed. To remove the worst radio frequency interference, any
frequency channel with power more than $4\sigma$ in excess of the
local median was also removed (``local'' was defined as the nearest 21
channels and the standard deviation $\sigma$ was determined
iteratively). CPSR2 also operated a total power monitor on microsecond
timescales, which removed most impulsive interference.

The CPSR2 data were next integrated for five minutes and calibrated
for differential gain and phase to correct for possible asymmetries in
the receiver hardware. If calibrator observations were available
(especially in the years directly following the CPSR2 commissioning,
observations of a pulsating noise source, needed for polarimetric
calibration, were not part of the standard observing
schedule). Subsequently the data were integrated for the duration of
the observation, which was typically 32\,minutes for PSRs
J2124$-$3358, J1939+2134 and J1857+0943 and 64\,minutes for all other
pulsars. In the case of PSR J1643$-$1224, the integration time was
32\,minutes until December 2005 and 64\,minutes from 2006
onwards. Finally, the CPSR2 data were integrated in frequency and the
Stokes parameters were combined into total power. CPSR2 data that did
not have calibrator observations available were processed identically,
except for the calibration step. While for some pulsars (like PSR
J0437$-$4715) these uncalibrated data are provably of inferior quality
\citep[see, e.g.][]{van06}, in our case this is largely outweighed by
the improved statistics of the larger number of TOAs and by the
extended timing baseline these observations provided. We therefore
include both calibrated and uncalibrated observations in our data
sets.

Pulse TOAs were determined through cross-correlation of the total
intensity profiles thus obtained with pulsar and frequency-dependent
template profiles. These template profiles were created through
addition of a large number of observations and were phase-aligned for
both CPSR2 observing bands. As there were only few high
signal-to-noise observations obtained with the FPTM and FB backends
for most pulsars, these data were timed against standards created with
the CPSR2 backend.  This may affect the reliability of their derived
TOA errors. For this reason we have evaluated the underestimation of
TOA errors for each backend separately, as explained in the next
section. While the TOA errors were generally determined through the
standard Fourier phase gradient method, the Gaussian interpolation
method produced more accurate estimates for pulsars with low
signal-to-noise ratios \citep{hbo05a} - specifically for PSRs
J0613$-$0200, J2129$-$5721, J1732$-$5049, J2124$-$3358 and
J1045$-$4509. The PSRCHIVE software package \citep{hvm04} was used to
perform all of the processing described above.
\begin{table*}
	\begin{minipage}{17.5cm}
      \caption{Summary of the timing results, sorted in order of
        decreasing rms residual. The columns present the pulsar name,
        the rms timing residual (without prewhitening), the length of
        the data set and the number of TOAs. For PSRs J1939+2134 and
        J1857+0943 this table only contains the Parkes data. See
        \S\ref{sec:timing} and \S\ref{sec:stab} for
        details.}\label{tab:summary}
	  \begin{tabular}{c|r@{.}l|r@{.}l|r}
	    \hline Pulsar & \multicolumn{2}{c}{rms} & \multicolumn{2}{c}{T}
      & \multicolumn{1}{c}{N$_{\rm TOA}$}\\
      name & \multicolumn{2}{c}{($\mu$s)} & \multicolumn{2}{c}{(yr)}& 
      \\
	    \hline 
	    J1909$-$3744 & 0&166 & 5&2  & 893 \\ 
	    J1713+0747   & 0&198 & 14&0 & 392 \\ 
	    J0437$-$4715 & 0&199 & 9&9  &2847 \\ 
	    J1744$-$1134 & 0&617 & 13&2 & 342 \\ 
	    J1939+2134   & 0&679 & 12&5 & 168 \\ 
	    \\                               
	    J1600$-$3053 & 1&12  & 6&8  & 477 \\ 
	    J0613$-$0200 & 1&52  & 8&2  & 190 \\ 
	    J1824$-$2452 & 1&63  & 2&8  &  89 \\ 
	    J1022+1001   & 1&63  & 5&1  & 260 \\ 
	    J2145$-$0750 & 1&88  &13&8  & 377 \\ 
	    \\                               
	    J1643$-$1224 & 1&94  &14&0  & 241 \\ 
	    J1603$-$7202 & 1&98  &12&4  & 212 \\ 
	    J2129$-$5721 & 2&20  &12&5  & 179 \\ 
	    J1730$-$2304 & 2&52  &14&0  & 180 \\ 
	    J1857+0943   & 2&92  & 3&9  & 106 \\ 
	    \\                               
	    J1732$-$5049 & 3&23  & 6&8  & 129 \\ 
	    J0711$-$6830 & 3&24  &14&2  & 227 \\ 
	    J2124$-$3358 & 4&01  &13&8  & 416 \\ 
	    J1024$-$0719 & 4&17  &12&1  & 269 \\ 
	    J1045$-$4509 & 6&70  &14&1  & 401 \\ 
	    \hline
	  \end{tabular}
  \end{minipage}
\end{table*}
\begin{landscape}
\pagestyle{empty}
\oddsidemargin=-1.4cm
\topmargin=3cm
\begin{figure*}
  \centerline{\psfig{angle=-90.0,width=25.0cm,figure=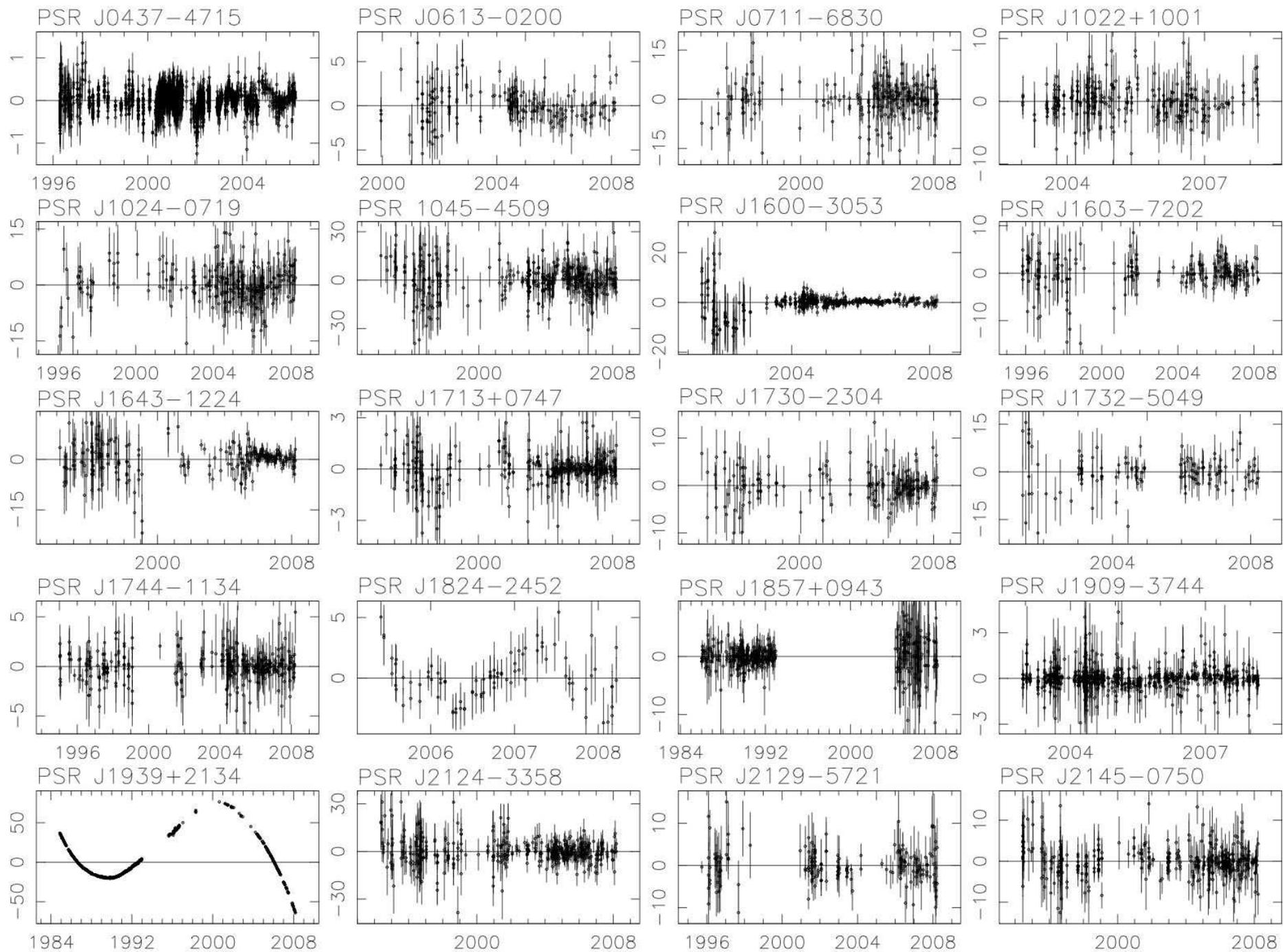}}
    \caption{Timing residuals of the 20 pulsars in our sample. Scaling
      on the x-axis is in years and on the y-axis in $\mu$s. For PSRs
      J1857+0943 and J1939+2134, these plots include the Arecibo data made
      publically available by Kaspi et al. (1994); all other data are from
      the Parkes telescope, as described in \S2. Sudden changes in white
      noise levels are due to changes in pulsar backend set-up - see \S2
      for more details.}
    \label{fig:Residuals}
\end{figure*}
\end{landscape}
\pagestyle{headings}
\begin{table*}
  \begin{center}
	\caption{Timing parameters for the single pulsars PSRs J0711$-$6830,
    J1024$-$0719, J1730$-$2304, J1744$-$1134, J1824$-$2452, J1939+2134
    and J2124$-$3358. Numbers in brackets give twice the formal
    standard deviation in the last digit quoted.  Note that these
    parameters are determined using \textsc{Tempo2}, which uses the
    International Celestial Reference System and Barycentric
    Coordinate Time. As a result this timing model must be modified
    before being used with an observing system that inputs Tempo
    format parameters. See \citet{hem06} for more
    information.}\label{tab:SinglePsrs}
	\begin{tabular}{lllll}
	  \hline
	  \multicolumn{5}{c}{Fit and data set parameters}\\
	  \hline
	  Pulsar name\dotfill                          & J0711$-$6830       & J1024$-$0719        &	J1730$-$2304     & J1744$-$1134      \\
	  \\                                                                                      
	  MJD range\dotfill                            & 49373.6$-$54546.4  & 50117.5$-$54544.6   & 49421.9$-$54544.8 & 49729.1$-$54546.9 \\
	  Number of TOAs\dotfill                       & 227                & 269                 & 180               & 342 \\
	  rms timing residual ($\mu$s)\dotfill         & 3.24               & 3.80                & 2.52              & 0.617 \\
	  Reference epoch for P, $\alpha$, \\                                                     
	  $\delta$ and DM determination\dotfill        & 49800              & 53000               & 53300             & 53742 \\
                                                 
	  \hline                                       
	  \multicolumn{5}{c}{Measured Quantities}\\    
	  \hline                                       
	  Right ascension, $\alpha$ (J2000.0)\dotfill  & 07:11:54.22579(15) & 10:24:38.68846(3)   & 17:30:21.6611(3)  & 17:44:29.403209(4) \\
	  Declination, $\delta$ (J2000.0)\dotfill      & $-$68:30:47.5989(7)& $-$07:19:19.1700(10)& $-$23:04:31.29(8) & $-$11:34:54.6606(2)\\
	  Proper motion in $\alpha$,$\mu_{\rm \alpha}  
    \cos{\delta}$(mas yr$^{-1}$)\dotfill         & $-$15.55(8)        & $-$35.3(2)           & 20.27(6)         & 18.804(15) \\
	  Proper motion in $\delta$, $\mu_{\rm \delta} $
	  (mas yr$^{-1}$)\dotfill                      & 14.23(7)           & $-$48.2(3)           & --               & $-$9.40(6) \\
	  Annual parallax, $\pi$ (mas)\dotfill         & --                 & --                   & --               & 2.4(2)  \\ 
	  Dispersion measure, DM (cm$^{-3}$ pc)\dotfill& 18.408(4)          & 6.486(3)             & 9.617(2)         & 3.1380(6) \\
	  Pulse frequency, $\nu$ (Hz)\dotfill          & 182.117234869347(4)& 193.71568356844(13) & 123.110287192301(2) & 245.4261197483027(5) \\
	  Pulse frequency derivative, 
    $\dot{\nu}$ ($10^{-16}\,$s$^{-2}$)\dotfill   & $-$4.94406(15)      & $-$6.95(3)        & $-$3.05906(10)      & $-$5.38188(4)\\
	  \hline 
    \multicolumn{5}{c}{Prewhitening Terms}\\
    \hline
    Fundamental wave frequency, $\omega_{\rm pw}$ (yr$^{-1}$)\dotfill
                                                 & --                 & 0.10368           & --                  & -- \\
    Amplitude of wave 1 cosine and sine terms,\\
    $A_{\cos,1}; A_{\sin,1}$ ($10^{-4}\,$s)\dotfill
                                                 & --                 & 2(13); 4.7(21)      & --                  & -- \\
    \hline
	  \\
	  \hline
	  \multicolumn{4}{c}{Fit and data set parameters}\\
	  \hline
	  Pulsar name\dotfill    & J1824$-$2452        & J1939+2134          & J2124$-$3358 \\
	  \\
	  MJD range\dotfill      & 53518.8$-$54544.9   & 46024.8$-$54526.9   & 49489.9$-$54528.9 \\
	  Number of TOAs\dotfill & 89                  & 180                 & 416 \\
	  rms timing residual 
    ($\mu$s)\dotfill       & 0.986               & 0.354               & 4.03 \\

	  Reference epoch for P, $\alpha$,\\
	  $\delta$ and DM determination\dotfill 
                           & 54219               & 52601               & 53174 \\
	  \hline
	  \multicolumn{4}{c}{Measured Quantities}\\
	  \hline
	  Right ascension, $\alpha$ (J2000.0)\dotfill 
                           & 18:24:32.00796(2)  & 19:39:38.561297(2)  & 21:24:43.85347(3) \\
	  Declination, $\delta$ (J2000.0)\dotfill 
                           & $-$24:52:10.824(6) & +21:34:59.12950(4) & $-$33:58:44.6667(7) \\
	  Proper motion in $\alpha$, $\mu_{\rm \alpha} \cos{\delta}$ (mas yr$^{-1}$)\dotfill 
                           &  --                & 0.072(2)             & $-$14.12(13) \\
	  Proper motion in $\delta$, $\mu_{\rm \delta}$ (mas yr$^{-1}$)\dotfill
                           &  $-$9(5)          & $-$0.415(3)          & $-$50.34(25) \\
	  Annual parallax, $\pi$ (mas)\dotfill 
                           & --                 & 0.13(13)              & 3.1(11) \\
	  Dispersion measure, DM (cm$^{-3}$ pc)\dotfill 
	                         & 120.502(2)         & 71.0227(9)          & 4.601(3) \\
	  Pulse frequency, $\nu$ (Hz)\dotfill  
	                         & 327.4055946921(6)& 641.928233642(12) & 202.793893879496(2) \\
	  Pulse frequency derivative, $\dot{\nu}$ ($10^{-16}\,$s$^{-2}$)\dotfill 
	                         & $-$1736.5(3)      & $-$429.1(6)      & $-$8.4597(2) \\
    \hline
    \multicolumn{4}{c}{Prewhitening Terms}\\
    \hline
    Fundamental wave frequency, $\omega_{\rm pw}$ (yr$^{-1}$)\dotfill
                            & 0.44734        & 0.14996         & -- \\
    Amplitudes of cosine and sine terms ($10^{-4}\,$s):\\
    wave 1: $A_{\cos,1}; A_{\sin,1}$\dotfill
                           & $-$20(6); 2.1(14) & 286(41); $-$413(60) & -- \\
    wave 2: $A_{\cos,2}; A_{\sin,2}$\dotfill
                           & --                 & 30(5); 84(12)     & -- \\
    wave 3: $A_{\cos,3}; A_{\sin,3}$\dotfill
                           & --                 & $-$21(3); $-$5.8(9)  & -- \\
    wave 4: $A_{\cos,4}; A_{\sin,4}$\dotfill
                           & --                 & 3.7(5); $-$2.9(5) & -- \\
    wave 5: $A_{\cos,5}; A_{\sin,5}$\dotfill
                           & --                 & 0.04(3); 0.68(9)& -- \\
	  \hline
	\end{tabular}
  \end{center}
\end{table*}
\clearpage 
\pagestyle{empty}
\begin{landscape}
\oddsidemargin=-1.4cm
\topmargin=6cm
  \begin{table*}
    \caption{Timing parameters for binary PSRs J1022+1001,
      J1600$-$3053, J1713+0747, J1857+0943, J1909$-$3744 and
      J2145$-$0750. See caption of Table 3 for more information.}
    \begin{tabular}{lllllll}
      \hline
      \multicolumn{7}{c}{Fit and data set parameters}\\
      \hline
      Pulsar name\dotfill & 
      J1022+1001          & J1600$-$3053          & J1713+0747           & J1857+0943          &  J1909$-$3744          & J2145$-$0750      \\
      \\
      MJD range\dotfill & 
      52649.7$-$54528.5   & 52055.7$-$54544.6     & 49421.9$-$54546.8    & 46436.7$-$54526.9   & 52618.4$-$54528.8      & 49517.8$-$54547.1 \\
      Number of TOAs\dotfill & 
      260                 & 477                   & 392                  & 376                 & 893                    & 377 \\
      rms timing residual ($\mu$s)\dotfill & 
      1.63                & 1.12                  & 0.198                & 1.14                & 0.166                  & 1.88 \\
      Reference epoch for P, $\alpha$, $\delta$ \\ 
      and DM determination\dotfill & 
      53589               & 53283                 & 54312                & 50481               & 53631                  & 53040 \\
      \hline
      \multicolumn{7}{c}{Measured Quantities}\\
      \hline
      Right ascension, $\alpha$ (J2000.0)\dotfill & 
      10:22:58.003(3)     & 16:00:51.903798(11)   & 17:13:49.532628(2)   & 18:57:36.392909(13) & 19:09:47.4366120(8)    & 21:45:50.46412(3) \\
      Declination, $\delta$ (J2000.0)\dotfill & 
      +10:01:52.76(13)    & $-$30:53:49.3407(5)   & +07:47:37.50165(6)   & +09:43:17.2754(3)   & $-$37:44:14.38013(3)   & $-$07:50:18.4399(14) \\
      Proper motion in $\alpha$, $\mu_{\rm \alpha} \cos{\delta}$ (mas yr$^{-1}$)\dotfill &
      $-$17.02(14)        & $-$1.06(9)            & 4.924(10)            & $-$2.64(3)          & $-$9.510(7)            & $-$9.66(15) \\
      Proper motion in $\delta$, $\mu_{\rm \delta}$ (mas yr$^{-1}$)\dotfill &
      --                  & $-$7.1(3)             & $-$3.85(2)           & $-$5.46(4)          & $-$35.859(19)          & $-$8.9(4) \\
      Annual parallax, $\pi$ (mas)\dotfill & 
      1.8(6)              & 0.2(3)                & 0.94(10)             & 1.1(4)              & 0.79(4)                & 1.6(5) \\
      Dispersion measure, DM (cm$^{-3}$ pc)\dotfill &
      10.261(2)           & 52.3262(10)           & 15.9915(2)           & 13.300(4)           & 10.3934(2)             & 8.9977(14) \\
      Pulse frequency, $\nu$ (Hz)\dotfill & 
      60.7794479762157(4) & 277.9377070984926(17) & 218.8118404414362(3) & 186.494078620232(2) & 339.31568740949071(10) & 62.2958878569665(6) \\
      Pulse frequency derivative, $\dot{\nu}$ ($10^{-16}$\,s$^{-2}$)\dotfill &
      $-$1.6012(2)        & $-$7.3390(5)          & $-$4.08379(3)        & $-$6.20495(6)       & $-$16.14819(5)         & $-$1.15588(3) \\
      \\
      Orbital period, $P_{\rm b}$ (days)\dotfill&
      7.8051302826(4)     & 14.3484577709(13)     & 67.825130963(17)     & 12.32719(4)         & 1.533449474590(6)      & 6.83893(2)\\
      Orbital period derivative, $\dot{P}_{\rm b}$ ($10^{-13}$)\dotfill &
      --                  & --                    & 41(20)               & 3(3)                & 5.5(3)                 & 4(3) \\
      Epoch of periastron passage, $T_{\rm 0}$ (MJD)\dotfill&
      53587.3140(6)       & 53281.191(4)          & 54303.6328(7)        & 50476.095(8)        & --                     & 53042.431(3) \\
      Projected semi-major axis, $x = a \sin{i}$ (s)\dotfill&
      16.7654074(4)       & 8.801652(10)          & 32.3424236(3)        & 9.230780(5)         & 1.89799106(7)          & 10.1641080(3) \\
      $\dot{x}$ ($10^{-14}$)\dotfill & 
      1.5(10)             & $-$0.4(4)             & --                   & --                  & $-$0.05(4)             & $-$0.3(3) \\
      Longitude of periastron, $\omega_{\rm 0}$ (deg)\dotfill&
      97.75(3)            & 181.85(10)            & 176.190(4)           & 276.5(2)            & --                     & 200.63(18) \\
      Orbital eccentricity, $e$ ($10^{-5}$)\dotfill & 
      9.700(4)            & 17.369(4)             & 7.4940(3)            & 2.170(6)            & --                     & 1.930(6) \\
      $\kappa = e \sin{\omega_{\rm 0}}$ ($10^{-8}$)\dotfill &
      --                  & --                    & --                   & --                  & $-$0.4(4)              & -- \\
      $\eta = e \cos{\omega_{\rm 0}}$ ($10^{-8}$)\dotfill & 
      --                  & --                    & --                   & --                  & $-$13(2)               & -- \\
      Ascending node passage, $T_{\rm asc}$ (MJD)\dotfill &
      --                  & --                    & --                   & --                  & 53630.723214894(4)     & -- \\
      Periastron advance, $\dot{\omega}$ (deg/yr)\dotfill & 
      --                  & --                    & --                   & --                  & --                     & 0.06(6) \\
      Sine of inclination angle, $\sin{i}$\dotfill&
      0.73\footnote{From Hotan et al. (2006).} & 0.8(4) & --             & 0.9990(7)           & 0.9980(2)              & -- \\
      Inclination angle, $i$ (deg)\dotfill & 
      47\addtocounter{mpfootnote}{-1}\mpfootnotemark{} & -- & 78.6(17)   & --                  & --                     & -- \\
      Companion mass, $M_{\rm c}$ ($M_{\odot}$)\dotfill & 
      1.05\addtocounter{mpfootnote}{-1}\mpfootnotemark{} & 0.6(15) & 0.20(2) & 0.27(3)        & 0.212(4)               & -- \\
      Longitude of ascending node, $\Omega$ (deg)\dotfill & 
      --                  & --                    & 67(17)               & --                  & --                     & -- \\
      \hline
    \end{tabular}
  \label{tab:binpsrs1}
\end{table*}
%
\begin{table*}
    \caption{Timing parameters for binary PSRs J0613$-$0200,
      J1045$-$4509, J1603$-$7202, J1643$-$1224, J1732$-$5049 and
      J2129$-$5721. See caption of Table 3 for more information.  }
    \begin{tabular}{lllllll}
      \hline
      \multicolumn{7}{c}{Fit and data set parameters}\\
      \hline
      Pulsar name\dotfill & 
      J0613$-$0200        & J1045$-$4509        &  J1603$-$7202       & J1643$-$1224       & J1732$-$5049         & J2129$-$5721     \\
      \\
      MJD range\dotfill & 
      51526.6$-$54527.3   & 49405.5$-$54544.5   & 50026.1$-$54544.7   & 49421.8$-$54544.7  & 52056.8$-$54544.8    & 49987.4$-$54547.1 \\
      Number of TOAs\dotfill &
      190                 & 401                 & 212                 & 241                & 129                  & 179 \\
      rms timing residual ($\mu$s)\dotfill  &  
      1.52                & 6.70                & 1.98                & 1.94               & 3.23                 & 2.20 \\
      Reference epoch for P, $\alpha$, $\delta$ \\
      and DM determination\dotfill & 
      53114               & 53050               & 53024               & 49524              & 53300                & 54000 \\
      \hline
      \multicolumn{7}{c}{Measured Quantities}\\
      \hline
      Right ascension, $\alpha$ (J2000.0)\dotfill & 
      06:13:43.975142(11) & 10:45:50.18951(5)   & 16:03:35.67980(4)   & 16:43:38.15544(8)  & 17:32:47.76686(4)    & 21:29:22.76533(5) \\
      Declination, $\delta$ (J2000.0)\dotfill & 
      $-$02:00:47.1737(4) & $-$45:09:54.1427(5) & $-$72:02:32.6985(3) & $-$12:24:58.735(5) & $-$50:49:00.1576(11) & $-$57:21:14.1981(4)\\
      Proper motion in $\alpha$, $\mu_{\rm \alpha} \cos{\delta}$ (mas yr$^{-1}$)\dotfill&
      1.84(8)             & $-$6.0(2)           & $-$2.52(6)          & 5.99(10)           & --                   & 9.35(10) \\
      Proper motion in $\delta$, $\mu_{\rm \delta}$ (mas yr$^{-1}$)\dotfill&
      $-$10.6(2)          & 5.3(2)              & $-$7.42(9)          & 4.1(4)             & $-$9.3(7)            & $-$9.47(10) \\
      Annual parallax, $\pi$ (mas)\dotfill & 
      0.8(7)              & 3.3(38)             & --                  & 2.2(7)             & --                   & 1.9(17) \\
      Dispersion measure, DM (cm$^{-3}$ pc)\dotfill & 
      38.782(4)           & 58.137(6)           & 38.060(2)           & 62.409(2)          & 56.822(6)            & 31.853(4) \\
      Pulse frequency, $\nu$ (Hz)\dotfill & 
      326.600562190182(4) & 133.793149594456(2) & 67.3765811408911(5) & 216.373337551614(7) & 188.233512265437(3) & 268.359227423608(3) \\
      Pulse frequency derivative, $\dot{\nu}$ ($10^{-16}$\,s$^{-2}$)\dotfill & 
      $-$10.2308(8)       & $-$3.1613(3)        & $-$0.70952(5)       & $-$8.6438(2)       & $-$5.0338(12)        & $-$15.0179(2) \\
      \\
      Orbital period, $P_{\rm b}$ (days)\dotfill&
      1.1985125753(1)     & 4.0835292547(9)     & 6.3086296703(7)     & 147.01739776(5)    & 5.262997206(13)      & 6.625493093(1) \\
      Epoch of periastron passage, $T_{\rm 0}$ (MJD)\dotfill&
      53113.98(2)         & 53048.98(2)         & --                  & 49577.9689(13)     & --                   & 53997.52(3) \\
      Projected semi-major axis, $x = a \sin{i}$ (s)\dotfill&
      1.0914444(3)        & 3.0151325(10)       & 6.8806610(4)        & 25.072614(2)       & 3.9828705(9)         & 3.5005674(7) \\
      $\dot{x}$ ($10^{-14}$)\dotfill & 
      --                  & --                  & 1.8(5)              & $-$4.9(5)          & --                   & 1.1(6) \\
      Longitude of periastron, $\omega_{\rm 0}$ (deg)\dotfill&
      54(6)               & 242.7(16)           & --                  & 321.850(3)         & --                   & 196.3(15) \\
      Orbital eccentricity, $e$ ($10^{-5}$)\dotfill & 
      0.55(6)             & 2.37(7)             & --                  & 50.578(4)          & --                   & 1.21(3) \\
      $\kappa = e \sin{\omega_{\rm 0}}$ ($10^{-6}$)\dotfill &
      --                  & --                  & 1.61(14)            & --                 & 2.20(5)              & -- \\
      $\eta = e \cos{\omega_{\rm 0}}$ ($10^{-6}$)\dotfill & 
      --                  & --                  & $-$9.41(13)         & --                 & $-$8.4(4)            & -- \\
      Ascending node passage, $T_{\rm asc}$ (MJD)\dotfill &
      --                  & --                  & 53309.3307830(1)    & --                 & 51396.366124(2)      & -- \\
      \hline
    \end{tabular}

  \label{tab:binpsrs2}
\end{table*}
\end{landscape}

\pagestyle{headings}
\subsection{Timing Results}\label{sec:timing}
The \textsc{tempo2} software package \citep{hem06} was used to
calculate the residuals from the TOAs and initial timing solutions
(Table \ref{tab:psrs}). In order to account for the unknown
instrumental delays and pulsar-dependent differences in observing
setup, arbitrary phase-offsets were introduced between the data from
different backends. Where available, data at an observing frequency of
685\,MHz were included in an initial fit to inspect visually for the
presence of dispersion measure (DM) variations. In the case of PSRs
J1045$-$4509, J1909$-$3744, J1939+2134 and J0437$-$4715, such
variations were significant and dealt with in the timing software
through a method similar to that presented by \cite{yhc+07}. The
average DM values presented in Tables \ref{tab:SinglePsrs},
\ref{tab:binpsrs1} and \ref{tab:binpsrs2} were determined from the
20\,cm data exclusively. The uncertainties in these DM values do not
take into account possible pulse shape differences between the
profiles at these slightly varying frequencies. We updated all the
pulsar ephemerides to use International Atomic Time (implemented as
TT(TAI) in \textsc{tempo2}) and the DE405 Solar System ephemeris
\citep{sta04b}.

In order to correct for any underestimation of TOA uncertainties
resulting from (amongst others) the application of CPSR2-based
template profiles to the FB and FPTM data (as mentioned in
\S\ref{ssec:ATD}) and to allow comparison of our timing model
parameters to those published elsewhere, the TOA uncertainties were
multiplied by error factors (so-called ``EFACs'') that are dependent
on the pulsar and observing system. Specifically, this part of the
analysis was performed as follows. First the timing data from each
observing system were prewhitened by fitting harmonically related
sine/cosine pairs if required. Next the TOA uncertainties were
multiplied by an EFAC value that produced a reduced $\chi^2$ value of
unity for that prewhitened subset of the data. Because of potential
non-Gaussian noise in the data, application of these backend-specific
EFACs does not necessarily result in a reduced $\chi^2$ value of unity
for the entire, recombined data set. To account for such non-Gaussian
noise in the data, a `global' EFAC was applied to the entire data set,
making the reduced $\chi^2$ after prewhitening equal to unity and
increasing the parameter uncertainties reported in the timing models
appropriately. As mentioned the prewhitening method was based on
fitting of sine/cosine pairs to the data, according to the following
formula described by \citet{mar01} \citep[and replicated in][]{hem06}:
\[
\Delta R = \sum_{k=1}^{n_{\rm H}} A_{\sin,k}\sin(k\omega_{\rm pw} \Delta t) 
+ A_{\cos,k}\cos(k\omega_{\rm pw} \Delta t)
\]
where $k$ runs over all sine/cosine pairs, $n_{\rm H}$ is the total
number of harmonically related pairs fitted, $A_{\cos,k}$ and
$A_{\sin,k}$ are respectively the amplitude of the $k^{\rm th}$ cosine
and sine waves and $\omega_{\rm pw}$ is the fundamental frequency
derived from:
\[
\omega_{\rm pw} = \frac{2 \pi}{T (1+4/n_{\rm H})}
\]
with $T$ the length of the data set. If prewhitening terms were
included in the final fit, we provide the values for $\omega_{\rm
  pw}$, $A_{\cos,k}$ and $A_{\sin,k}$ as part of our timing model.

Because the potential non-Gaussian noise present in the data is the
subject of our investigations in the remaining sections of this paper,
the prewhitening terms as well as the global EFACs were not included
in any subsequent analysis. The residuals plotted in Figure
\ref{fig:Residuals} and the parameters presented in Table
\ref{tab:summary} therefore do not include prewhitening terms or
global EFACs.

The system-specific EFACs were generally less than two, with the only
major outliers being the CPSR2 data of PSR J1939+2134 with an EFAC of
5.27 and the 32-min CPSR2 integrations (pre-2004 CPSR2 data) of
PSR J1643$-$1224, which have an EFAC of 4.9. In the former case this
large EFAC may be caused by incomplete prewhitening, as the
non-Gaussian noise is badly modelled by polynomials or sine/cosine
pairs. The underestimation of PSR J1643$-$1224 TOA uncertainties is
likely caused by the low signal-to-noise ratio of these observations,
which causes the Fourier phase gradient method to underestimate TOA
errors \citep[as previously reported by][]{hbo05a}. We note that the
EFAC for the 64\,minute integrations is much lower, at 2.5. In
deriving the timing models, the global EFAC was at most 1.1 and for
more than half of our sources less than 1.05.

The fact that most of our EFAC values are close to unity and show
little variation with backend, suggests that the parameter and error
estimates are fairly robust. In order to account for the different
sensitivity of the backends used and to limit effects of scintillation
on our timing, we opt for a weighted analysis. It is therefore
important to consider the impact of the TOA errors and applied EFACs
on the different parts of this analysis. Given that for most pulsars
the EFACs applied to the different backends are nearly equal, the
resulting timing models will be little affected by these EFAC
values. The reported uncertainties on the timing model parameters will
be affected but will be comparable to previous publications, since our
analysis method is similar. A full error analysis \citep[as suggested
  by][]{vbv+08} is needed to provide any more reliable parameter
uncertainties. Since the focus of this present paper is on the overall
timing stability and implications for pulsar timing array science, we
defer such error analysis (and the interpretation of any previously
unpublished parameters in our timing models) to a later paper. We
have, however, investigated the effect of weighting and EFACs on the
timing stability analysis (\S\ref{sec:stab}), but have not uncovered
any unexpected deviations beyond statistical noise. We therefore
conclude that the weighting and applied EFACs do not invalidate our
analysis.

A summary of the lengths of the data sets and the achieved rms
residual can be found in Table \ref{tab:summary}, highlighting the
superior residual rms of PSRs J1909$-$3744, J0437$-$4715 and
J1713+0747 when compared to other pulsars.  The timing residuals for
our data sets are presented in Figure \ref{fig:Residuals} and the
timing models are presented in Tables \ref{tab:SinglePsrs},
\ref{tab:binpsrs1} and \ref{tab:binpsrs2}, where $2\sigma$ errors are
given, in accordance with previous practice. We encourage observers to
use the improved models when observing. We also note that all but a
few of the parameters in our timing models are consistent with those
published previously.

\section{Pulsar Timing Stability}\label{sec:stab}
In \S\ref{sec:intro}, we demonstrated that one of two vital questions
relating to the potential of PTAs to detect a GWB is whether a low
residual rms can be maintained over long timespans (a property we
refer to as ``timing stability''). Effectively, this question breaks
down into two parts: to what degree of significance low-frequency
noise is present in our pulsar timing data and how any such
low-frequency noise can be expected to affect sensitivity to a GWB. In
order to answer this question fully, a spectral-analysis-based
investigation of pulsar timing residuals that includes identification
and modelling of potential non-Gaussian noise sources, would be
required. Because of various pulsar timing-specific issues such as
clustering of data, large gaps in data sampling and large variations
in error-bar size, however, standard spectral analysis methods fail to
provide reliable power spectra of pulsar timing data. We therefore use
the alternative approach provided by the $\sigma_{\rm z}$ statistic,
as described by \citet{mte97}. A brief explanation of this statistic,
along with a presentation of the $\sigma_{\rm z}$ values of our data
is presented in \S\ref{sec:sz} and a discussion of these results in
terms of PTA-science is provided in \S\ref{sec:stab:conc}.

\subsection{$\sigma_{\rm z}$ Stability Analysis}\label{sec:sz}
Originally proposed by \citet{mte97}, the $\sigma_{\rm z}$ statistic
is defined as:
\[
\sigma_{\rm z}(\tau) = \frac{\tau^2}{2 \sqrt{5}} \langle
c_3^2\rangle^{1/2},
\]
where $\langle\rangle$ denotes the average over subsets of the data,
$c_3$ is determined from a fit of the polynomial
\[
c_0 + c_1 (t-t_0) +c_2(t-t_0)^2+c_3(t-t_0)^3
\] 
to the timing residuals for each subset and $\tau$ is the length of
the subsets of the data. In order for the $\sigma_{\rm z}$ values to
be independent of each other, we use $\tau = T, T/2, T/4, T/8,$ \ldots
only. The interpretation of this statistic in terms of power spectra
deserves some attention. As presented by \citet{mte97}, a power
spectrum with spectral index $\beta$:
\[
P(\nu) \propto f^{\beta}
\]
would translate into a $\sigma_{\rm z}$ curve:
\[
\sigma_{\rm z}(\tau) \propto \tau^{\mu},
\]
where the spectral indices are related as:
\begin{equation}\label{eq:SpindxRel}
  \mu = 
  \begin{cases}
	-\frac{1}{2}(\beta+3) & {\rm if\ } \beta < 1\\
	-2         & {\rm otherwise.}
  \end{cases}
\end{equation}

\begin{figure*}
  \includegraphics[width=17cm,angle=0.0]{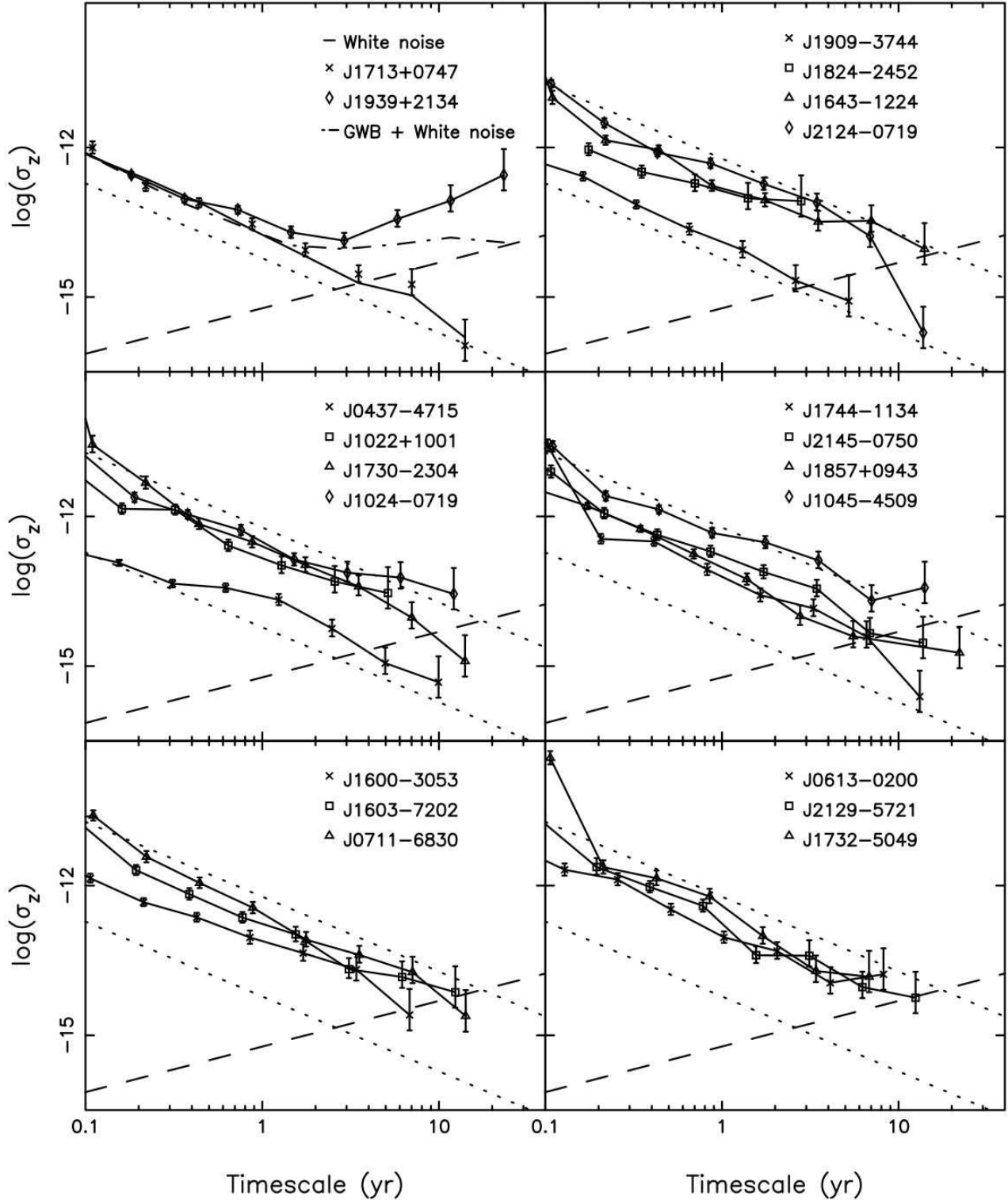}
  \caption{$\sigma_{\rm z}$ stability parameter for the 20 pulsars in
    our sample. The dotted slanted lines represent white noise levels
    of 100\,ns (bottom) and 10\,$\mu$s (top); the dashed slanted line
    shows the steepness introduced to pre-fit residuals by a
    hypothetical GWB (see \S\ref{sec:intro}); pulsars whose curve is
    steeper than this line (like PSR J1939+2134), can therefore be
    expected to be of little use to PTA efforts on long
    timescales. The top left figure further shows the average
    $\sigma_{\rm z}$ values resulting from 1000 simulations of white
    noise residuals sampled at the times of the PSR J1713+0747 data
    set and fitted for the PSR J1713+0747 timing model parameters
    (full line). This demonstrates that the PSR J1713+0747 data do not
    - within the sensitivity provided by the $\sigma_{\rm z}$
    statistic - contain a significant, steep red-noise process. The
    dash-dotted line in the top left figure shows the average of 2000
    simulations for white noise combined with a GWB, sampled at the
    times of the PSR J1939+2134 data set and fitted for pulse period
    and period derivative. These simulated results provide an example
    of the combined effect sampling and model fitting can have on the
    $\sigma_{\rm z}$ statistic, even in the case of white noise.}
  \label{fig:sigmaz}
\end{figure*}

Equation (\ref{eq:SpindxRel}) implies that spectra have different
slopes in a $\sigma_{\rm z}$ graph than in a power spectrum. Along
with the $\sigma_{\rm z}$ graphs for our data sets, Figure
\ref{fig:sigmaz} provides some examples of spectra for guidance: lines
with a slope of $-3/2$ (dotted lines in Figure \ref{fig:sigmaz})
represent spectrally white data ($\beta = 0$ into Equation
(\ref{eq:SpindxRel}) gives $\mu = -3/2$) and a GWB with a spectral
index $\alpha = -2/3$ in the gravitational strain spectrum (and
therefore a spectral slope $\beta = -13/3$ in the timing residual
spectrum, as follows from equations (\ref{eq:GWBPower}) and
(\ref{eq:h_c})) would have a positive slope of $2/3$ in $\sigma_{\rm
  z}$ (dashed lines).

Comparison of such theoretical slopes to the data is, however,
non-trivial since the data are strongly affected by effects from
sampling and fitting. As an illustration of such effects, the top-left
plot of Figure \ref{fig:sigmaz} shows two $\sigma_{\rm z}$ curves
derived from simulations. The first one is the full line that
approximates the $\sigma_{\rm z}$ curve for PSR J1713+0747. In this
case the $\sigma_{\rm z}$ values of 1000 simulations of white noise
with the timing rms and sampling of the PSR J1713+0747 data set were
averaged. This curve is not perfectly parallel to the theoretical
curve with slope $-3/2$ due to sampling, varying TOA uncertainties and
model fitting. Comparison of the white noise simulations with the
actual PSR J1713+0747 data indicates that there is not a significant,
steep red-noise process affecting the timing residuals for this
pulsar. The second simulation in the top-left plot of Figure
\ref{fig:sigmaz} is the dot-dashed line, which is the average
$\sigma_{\rm z}$ graph of 2000 simulations of white noise with an
artificial GWB and the sampling of the PSR J1939+2134 data set, fitted
for pulse period and spindown. This simulated curve does not reach the
theoretical slope of $2/3$ because of flattening off at low
frequencies caused by sampling, fitting and leakage resulting from
these. This simulation also demonstrates that the PSR J1939+2134 curve
is significantly steeper than a simulated GWB, implying that this
pulsar will most likely not be very useful for long-term PTA projects,
although its low rms residual on short time spans might make it useful
for detection of burst-type sources.

\begin{landscape}
\pagestyle{empty}
\oddsidemargin=-1.4cm
\topmargin=3cm
\begin{figure*}
  \centerline{\psfig{angle=0.0,height=18.0cm,figure=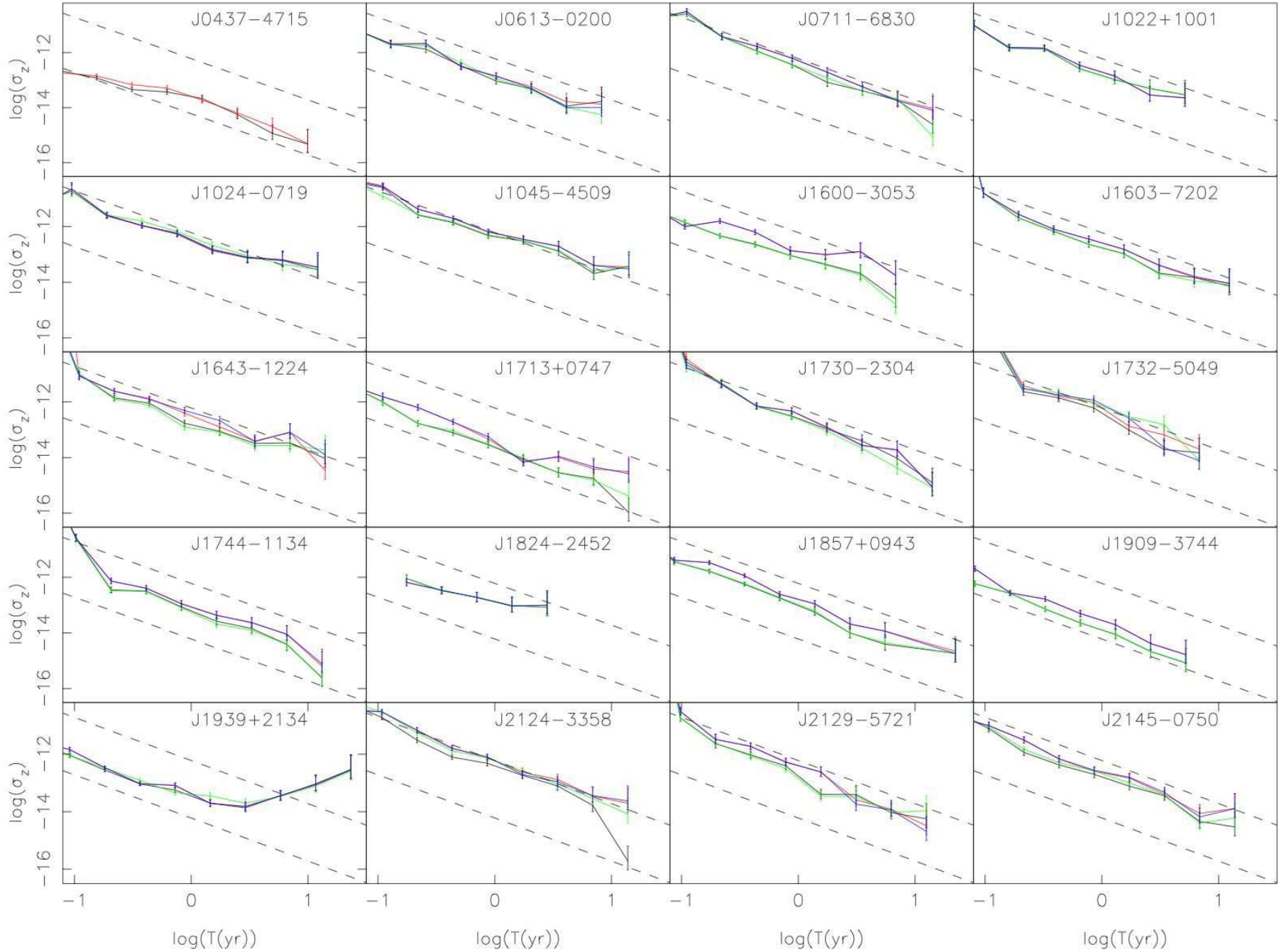}}
  \caption{$\sigma_{\rm z}$ graphs for all pulsars in our sample,
    showing both weighted (black and green lines) and unweighted (red
    and blue lines) results, as well as results including EFACs (black
    and red lines) and excluding EFACs (green and blue lines). The
    dashed lines represent theoretical white noise at levels of
    100\,ns and 10\,$\mu$s. The EFAC values of the PSR J0437$-$4715
    data were lost in processing, so the green and blue curves are
    missing for that particular pulsar. For all other pulsars all four
    curves are present, though they do frequently overlap.}
  \label{fig:SigWgt} 
\end{figure*}
\end{landscape}

\pagestyle{headings}
The $\sigma_{\rm z}$ graphs of our data are shown in Figure
\ref{fig:sigmaz}. A comparison of these curves to those obtained from
an unweighted analysis or from an analysis that does not contain the
EFAC values described in \S\ref{sec:timing}, is presented in Figure
\ref{fig:SigWgt}, a colour version of which is available online. This
graph demonstrates that use of weighting or EFACs does not affect the
data in any statistically significant manner, other than to decrease
the effect of the white noise component in case of scintillating
pulsars.

Comparison of the PSR J1713+0747 and PSR J1939+2134 data with the
simulated curves shown, along with the invariability of the shape of
$\sigma_{\rm z}$ plots to weighting or application of error factors,
shows that the $\sigma_{\rm z}$ parameter provides a good first-order
discrimination between pulsars that do not exhibit significant, steep
red noise (like PSR J1713+0747) and those that do have timing
instabilities which could mask a GWB (like PSR J1939+2134).

\subsection{Timing Stability Conclusions}\label{sec:stab:conc}

Figure \ref{fig:sigmaz} shows that PSR J1939+2134 has red noise with a
level and steepness that will limit its use in GWB-detection efforts
that last more than about two years. Four other pulsars (PSRs
J0613$-$0200, J1024$-$0719, J1045$-$4509 and J1824$-$2452) show some
indication of similar red noise, but longer timing and lower
white-noise levels are needed to determine this with statistical
significance. For all the other pulsars we have no evidence that the
red noise that may be present in the timing residuals below the
white-noise level has a spectral index that prevents GWB detection on
timescales of five years to a decade.

We have been unable to detect timing instabilities with an amplitude
and spectral slope that could mask a GWB in the timing data of PSRs
J1713+0747 and J1744$-$1134, notwithstanding their long data spans and
low timing rms which should make them highly sensitive to any
low-frequency noise. Using Equation (\ref{eq:ScalingLaw}), it can be
shown that the data sets of PSRs J1713+0747 and J0437$-$4715 already
meet the requirements for a ten-year long PTA experiment, proving that
at least for some pulsars the timing stability and rms residual
required to detect a GWB on timescales of ten years or more is
achievable. The challenge for such long-term projects will therefore
be to find more pulsars like these, or to replicate these results for
other existing pulsars, by increasing the sensitivity of observing
systems.

It must be noted that the study of irregularities in pulsar timing
data (often referred to as ``timing noise'') can be much more
extensive than presented here. Given our main aim of assessing the
impact on PTA science and the absence of clear timing noise in most of
our data sets, precise modelling or bounding of timing irregularities
as well as a thorough discussion of the potential sources of any
observed timing instabilities, has not been included in this analysis.

\section{Analysis of Residual rms}\label{sec:stability}
As an alternative to the long-term PTA detection efforts discussed in
the previous section, a shorter-term detection is possible if timing
at lower residual rms is achievable. The standard scenario for a
(relatively) short-term GWB detection by a PTA requires five years of
weekly observations with a timing rms of 100\,ns for 20 MSPs
\citep{jhlm05}. Since a residual rms of 100\,ns has never been
maintained over five years, the possibility that some intrinsic
property of MSPs induces instabilities at that level, remains open. In
this section, we will address that issue by evaluating how much the
timing rms of some of our most precisely timed pulsars may be reduced.

We separate three different categories of contributions to the pulsar
timing residuals:
\begin{description}
  \item{\bf Radiometer noise $\sigma_{\rm Rad}$:} The Gaussian noise
    component that scales with the radiometer equation and which is
    mainly determined by the shape and signal-to-noise (S/N) ratio of
    the observed pulse profiles.
  \item{\bf Frequency-systematic effects $\sigma_{\nu}$:} This
    category of noise contributions contains most effects that produce
    timing residuals dependent on the observing frequency. This
    includes interstellar effects such as interstellar scintillation
    and DM variations.
  \item{\bf Temporal-systematic effects $\sigma_{\tau}$:} This
    category contains all time-dependent effects such as calibration
    errors, instabilities in the observing systems, clock errors,
    errors in the Solar System ephemerides, GWs and intrinsic pulsar
    timing noise.
\end{description}

As it is impossible to get direct measures of the three contributions
listed above, we base our analysis on the following three
measurements:
\begin{description}
  \item{\bf Total timing rms $\sigma$:} This is simply the timing
    residual rms of the data considered. It contains all three
    effects:
    \begin{equation}\label{eq:Totrms}
      \sigma^2 = \sigma_{\rm Rad}^2 + \sigma_{\nu}^2 + \sigma_{\tau}^2.
    \end{equation}
  \item{\bf Sub-band rms $\sigma_{\rm sb}$:} In \S\ref{sec:ISM}, we
    will introduce this new measure which is $1/\sqrt{2}$ times the
    weighted rms of the offset between the residuals of two
    simultaneous observing bands with different centre
    frequencies. Since the observations in the two observing bands are
    simultaneous, their offset is determined by the radiometer noise
    and by frequency-systematic effects (as the observing bands are
    centred at slightly different frequencies). We can therefore
    write:
    \begin{equation}\label{eq:Sbrms}
      \sigma_{\rm sb}^2 = \sigma_{\rm Rad}^2 + \sigma_{\nu}^2.
    \end{equation}
  \item{\bf Theoretical radiometer noise $\sigma_{\rm Rad}$:} In
    \S\ref{sec:theory}, we will calculate $\sigma_{\rm Rad}$ directly
    from the pulse profiles used in our timing.
\end{description}

Using these three measures and equations (\ref{eq:Totrms}) and
(\ref{eq:Sbrms}), the three contributions to the timing residuals can be
isolated, the results of which are described in \S\ref{sec:stab:disc}.

Our analysis will be based on the CPSR2 data of PSRs J1909$-$3744,
J1713+0747 and J1939+2134. We restrict this analysis to the CPSR2 data
because it is of superior quality to the data of older backend systems
(see \S\ref{sec:obsSys}) and because it consists of the five most
densely sampled years of observations. We focus on three of the most
precisely timed pulsars in order to obtain the best limits on
achievable residual rms. In doing so, we omit PSR J0437$-$4715 because
the advanced calibration schemes used in its analysis
\citep[see][]{vbv+08,van04a,van06} complicates our efforts and because
reported non-Gaussian noise in the timing data of this pulsar
\citep{vbv+08} may imply an inferior limit to that derived from PSRs
J1909$-$3744 and J1713+0747. Note that the purpose of this analysis is
to uncover the \emph{potential} limit for high-precision timing: it is
already known (see e.g. \S\ref{sec:stab}) that MSPs have different
amounts of time-dependent noise, so the limit we will derive from PSRs
J1909$-$3744 and J1713+0747 does not have to hold for all
MSPs. However, it does suggest that other pulsars may achieve similar
rms residual and that a PTA-size sample of 20 MSPs at such rms
residual may mainly depend on increased sensitivity of current
observing systems and new discoveries in ongoing and future surveys.

\begin{table*}
  \begin{center}
	\caption{Breakdown of weighted timing residuals for three selected
      pulsars. Given are the total timing rms of the $\sim$5 years of
      CPSR2 data ($\sigma$), the sub-band timing rms ($\sigma_{\rm
        sb}$), the radiometer noise ($\sigma_{\rm Rad}$), the temporal
      systematic ($\sigma_{\tau}$) and the frequency systematic
      ($\sigma_{\nu}$) contributions to the timing rms. All values are
      in ns and apply to 64\,min integrations. See
      \S\ref{sec:stability} for more information.}
	\label{tab:stability}
	\begin{tabular}{c|c|c|c|c|c}
	  \hline
	  Pulsar name & $\sigma$ & $\sigma_{\rm sb}$ & $\sigma_{\rm Rad}$
	  & $\sigma_{\tau}$ & $\sigma_{\nu}$\\
    (1) & (2) & (3) & (4) & (5) & (6) \\
	  \hline
	  J1909$-$3744 & 166 & 144 & 131 & 83 & 60 \\
	  J1713+0747   & 170 & 149 & 105 & 82 & 106 \\
	  J1939+2134   & 283 & 124 & \hspace{1ex}64 & 254 & 106 \\
	  \hline
	\end{tabular}
  \end{center}
\end{table*}
\subsection{Theoretical Estimation of Radiometer Noise, $\sigma_{\rm Rad}$}
\label{sec:theory}
The level at which the radiometer noise adds to the timing residuals
can be determined based on the pulsar's observed pulse profile shape
and brightness, as described by \citet{van06}. Equation (13) of that
publication provides the following measure (notice we only consider
the total intensity, $S_{\rm 0}$, to allow direct comparison with our
timing results):
\begin{equation}\label{eq:RadiometerNoise}
\sigma_{\rm Rad} = P\times \sqrt{V} = P\times \Bigg(4\pi^2
	\sum_{m=1}^{N_{\rm max}\leq N/2} \nu_{\rm m}^2 \frac{S_{\rm
	0,m}^2}{\varsigma_{\rm 0}^2}\Bigg)^{-0.5},
\end{equation}
where $\nu_{\rm m}$ is the $m^{\rm th}$ frequency of the Fourier
transform of the pulse profile, $S_{\rm 0,m}^2$ is the total power at
that frequency, $\varsigma_{\rm 0}$ is the white noise variance of the
profile under consideration, $N$ is the total number of time bins
across the profile and $N_{\rm max}$ is the frequency bin where the
Fourier transform of the pulse profile reaches the white noise level,
$\varsigma_{\rm 0}$. $V$ is the expected variance in the phase-offset
or residual, $P$ is the pulse period and $\sigma_{\rm Rad}$ is the
residual rms predicted for the pulse profile considered.

In order to use Equation (\ref{eq:RadiometerNoise}) on our data, we
first integrated all our pulse profiles together, weighted by
signal-to-noise ratio, after which Equation (\ref{eq:RadiometerNoise})
was applied to the final profile. Subsequently $\sigma_{\rm Rad}$ was
renormalised to 64\,min integrations through use of the radiometer
equation. In order to check this result, we
also applied the equation to all individual pulse profiles contained
in this analysis and averaged the results in a weighted way -
resulting in the same answer, which is given in column four of Table
\ref{tab:stability}. The value for PSR J1909$-$3744 shows that even at
this low residual rms, radiometer noise dominates the timing
rms. Applying this method to the other MSPs in our sample, we found
that almost all our timing residuals are dominated by radiometer
noise. For more than half of our sample of 20 MSPs, $\sigma_{\rm Rad}$
is of the order of a microsecond or more. This demonstrates the need
for longer integration times, larger bandwidth and/or larger
collecting area.

\subsection{Estimating Frequency-Dependent Effects}\label{sec:ISM}
As described in \S\ref{sec:obsSys}, the CPSR2 pulsar backend records
two adjacent, 64\,MHz-wide frequency bands simultaneously. This allows
determination of a unique measure of a sub-set of timing
irregularities, which we will call the ``sub-band rms'', $\sigma_{\rm
  sb}$:
\begin{equation}\label{eq:SubBandrms}
  \sigma_{\rm sb} = \frac{1}{\sqrt{2}}
  \sqrt{
    \frac{
      \sum_{i}{
        \frac{\left(r_{i,{\rm m}}-r_{i,{\rm n}}\right)^2}
             {e_{i,{\rm mn}}^2}
      }
    }
    {\sum_{i}{1/e_{i,{\rm mn}}^2}}
  },
\end{equation}
where the sums run over all observing epochs $i$, $r_{i,{\rm m}}$ and
$r_{i,{\rm n}}$ are the residuals of either observing band (named m
and n respectively) at epoch $i$ and $e_{i,{\rm mn}} = \sqrt{e_{i,{\rm
      m}}^2+e_{i,{\rm n}}^2}$ is the average TOA error at epoch
$i$. Effectively, the sub-band rms is $1/\sqrt{2}$ times the weighted
rms of the offset between the residuals of the two bands. This implies
it contains all contributions to the total rms that are not
time-dependent but either statistically white or dependent on the
observing frequency, as described earlier. Note, however, that many of
these effects have both a temporal and frequency component. Given our
sampling, it should therefore be understood that (specifically in the
case of DM variations) only part of these effects is contained in
$\sigma_{\nu}$, while the remaining contributions are contained in
$\sigma_{\tau}$.

The sub-band rms for the three selected pulsars is presented in column
three of Table \ref{tab:stability}.

\subsection{Discussion}\label{sec:stab:disc}
Based on equations (\ref{eq:Totrms}) and (\ref{eq:Sbrms}) and the
three measures $\sigma$, $\sigma_{\rm sb}$ and $\sigma_{\rm Rad}$
determined in the preceding paragraphs, the three contributions to the
rms ($\sigma_{\rm Rad}$, $\sigma_{\nu}$ and $\sigma_{\tau}$) can now
be estimated. Their values are presented in columns 4, 5 and 6 of
Table \ref{tab:stability}.  In order to assess the potential for
100\,ns timing of these sources over a five-year timescale, we will
now discuss the possible means of reducing these three contributions.

The radiometer noise $\sigma_{\rm Rad}$ scales for different
telescopes or observing systems according to the radiometer equation:
\begin{equation}\label{eq:Radiom}
  \sigma_{\rm Rad} \propto \frac{T_{\rm sys}}{ A_{\rm eff}\sqrt{B t}}
\end{equation}
where $B$ is the bandwidth of the observing system used, $t$ is the
integration time, $A_{\rm eff} = \eta \frac{\pi D^2}{4}$ is the
effective collecting area of the telescope (with $\eta$ the aperture
efficiency and $D$ the telescope diameter) and $T_{\rm sys}$ is the
system temperature of the receiver.

The frequency systematic contributions are not as easily scaled for
different observing systems, but they can be decreased and research on
this front is progressing \citep{yhc+07,hs08,wksv08}. Also, by
reducing the radiometer noise, any measurements of DM variations will
become more precise, which will enhance corrections for these effects
and therefore decrease the contribution of $\sigma_{\nu}$. We also
note that since these effects are frequency dependent, the employment
of very large bandwidth receivers or coaxial receiver systems such as
the 10/50\,cm receiver at the Parkes observatory, may lead to highly
precise determination and correction of these effects. Furthermore,
increased collecting area and bandwidth may enable future timing
observations at higher observational frequencies, which would limit
the size of these effects. We therefore suggest that $\sigma_{\nu}$
does not ultimately limit the achievable rms residual, but may largely
be corrected for if current research and technological development
progress.

The wide variety of sources that add to the temporal systematic make
predictions about its future evolution hard. Sources such as intrinsic
pulsar timing noise are (as yet) impossible to mitigate. Errors in the
terrestrial clocks or in the Solar System ephemerides are expected to
decrease as better models become available or as timing arrays provide
their own improved solutions for these models. Instabilities in the
observing system may to some degree be mitigated by improved
calibration methods \citep{van04a,van06}. Simultaneous observations of
a single source at multiple observatories may also lead to detection
and correction of instrumental instabilities and the time-dependent
effect of DM variations may also be mitigated, as explained above. 

Following from the above, we stress the fact that all contributions to
$\sigma_{\nu}$ and $\sigma_{\rm Rad}$ may be mitigated, but that
certain contributions to $\sigma_{\tau}$ cannot be corrected. This
implies that this last class of effects will ultimately limit the
residual rms that can be reached. We will therefore use the temporal
systematic contribution to the rms ($\sigma_{\tau}$, column five in
Table \ref{tab:stability}) as an upper limit on the potential rms
residual of the MSPs under investigation. Note that this is a
conservative upper limit since significant protions of $\sigma_{\tau}$
may be expected to be mitigated. However, without relative
quantification of the various contributions to $\sigma_{\tau}$, this
limit cannot be reliably decreased.

Given the discussion above, we note that the potential timing residual
rms of PSRs J1909$-$3744 and J1713+0747 is predicted to be below
100\,ns on a five-year timescale. This implies that the standard
scenario of 100\,ns timing over five years is possible provided
techniques currently being developed for mitigation of
frequency-dependent effects are successful, more sensitive observing
systems are used and more bright, stable MSPs like PSRs J1909$-$3744
and J1713+0747 are found.

\section{Prospects for Gravitational Wave Detection}
\label{sec:PTA}
\citet{jhlm05} derived the expected sensitivity of a PTA to a GWB with
given amplitude, $A$, both for homogeneous arrays (where all pulsars
have comparable timing residuals) and inhomogeneous arrays. They also
pointed out the importance of prewhitening\footnote{In this context,
  prewhitening refers to a technique that flattens the power spectrum
  of a time series by means of weighting. This flattening optimises
  the sensitivity of a PTA to steep red spectra such as those
  introduced by a GWB.} the residuals to increase sensitivity at
larger GWB amplitudes. In the current section, we will build upon
their analysis to provide more realistic predictions for ongoing and
future timing arrays. We extend their analysis in three fundamental
ways.

Firstly, we use the rms timing residuals presented in Table
\ref{tab:summary}. These results provide an inhomogeneous set of rms's
with a realistic spread. We assume the residuals are statistically
white and will therefore not change with the timescale of the timing
array project. Our analysis in \S\ref{sec:stab} shows that for most
pulsars this assumption is reasonable, especially on timescales of
order five years.

Secondly, we do not apply exactly the same algorithm as
\citet{jhlm05}. In Appendix \ref{app:Sens}, we present a derivation of
PTA sensitivity to a GWB in a manner that provides some guidance on
analysing the data. We assume that the prewhitening and correlation
are handled together by computing cross-spectra and we estimate the
amplitude of the GWB directly rather than using the normalised cross
correlation function. We assume that the non-GWB noise is white, but
can be different for each pulsar. Our results are very close to those
of \citet{jhlm05} and using our method we successfully reproduced the
scaling law, Equation (\ref{eq:ScalingLaw}). The analysis could be
easily extended to include non-white noise if a model for the noise
were available.

Finally, in order to generalise the results from our Parkes data to
telescopes in other parts of the world, we scale the residuals based
on realistic parameters for various PTA efforts listed in Table
\ref{tab:PTAs}. In doing so, we scale $\sigma_{\rm Rad}$ (see
\S\ref{sec:stability}) according to Equation (\ref{eq:Radiom}). As
discussed in \S\ref{sec:stab:disc}, some improvements in
$\sigma_{\nu}$ and $\sigma_{\tau}$ can be expected in coming years,
especially as the radiometer noise is decreased. While quantification
of any such improvement is practically impossible, we will apply the
same radiometer scaling to $\sigma_{\nu}$ as we apply to $\sigma_{\rm
  Rad}$ and assume $\sigma_{\tau}$ to be constant at $80\,$ns for all
pulsars at all telescopes. This may provide a slight disadvantage for
larger telescopes, but overall we consider this a reasonable yet
conservative approach.

\subsection{Ongoing PTA Projects}\label{sec:PTAs}
We consider five ongoing PTA efforts: 
\begin{description}
\item{\bf Current:} Refers to the data presented in this paper, using
  the longest overlapping time span of the sample: five years. This
  ignores the fact that the PSR J1824$-$2452 data set is shorter, but
  this globular cluster pulsar may not prove useful in a PTA project
  lasting longer than a few years anyway. We therefore assume that a
  replacement is found and has identical timing rms over a time span
  of five years.
\item{\bf Predicted PPTA:} Assumes the usage of 256\,MHz of bandwidth
  at the Parkes telescope, which implies a four-fold bandwidth
  increase and therefore a two-fold decrease in timing rms. The
  PPTA is the only one to be considered for more than five years,
  mainly in order to demonstrate the large impact a doubling of
  campaign length can have, but also because several years of high
  precision timing data with that bandwidth do already exist
  \citep{man08} for all 20\,MSPs.
\item{\bf NANOGrav:} Assumes Arecibo gain for the ten least well-timed
  pulsars and GBT gain for the ten best-timed pulsars, in order to get
  a fairly equal rms for all 20 MSPs. (Since we consider
  $\sigma_{\tau}$ an upper limit on the rms residual, the advantage
  of Arecibo over the GBT is limited for the brightest sources.)
\item{\bf EPTA:} Assumes monthly observations with five 100\,m-class
  telescopes \citep{jsk+08}.
\item{{\bf EPTA--LEAP}\footnote{Large European Array for Pulsars}{\bf:}}
  Interferometrically combines the five telescopes of the EPTA to form
  a single, larger one. This decreases the number of observations, but
  increases the gain.
\end{description}

An important caveat to this analysis is that several of the pulsars
under consideration cannot be observed with most Northern telescopes,
because of the telescope declination limits. We therefore assume
stable MSPs to be discovered in the Northern hemisphere. As mentioned
before, we also assume that progress will be made in the mitigation of
frequency-dependent ISM and calibration effects. Finally, this
analysis is based on the Parkes data presented in this paper and
therefore assumes systematic effects to be at most at the level of the
Parkes observing system used.

\begin{figure*}
  \includegraphics[width=10cm,angle=-90.0]{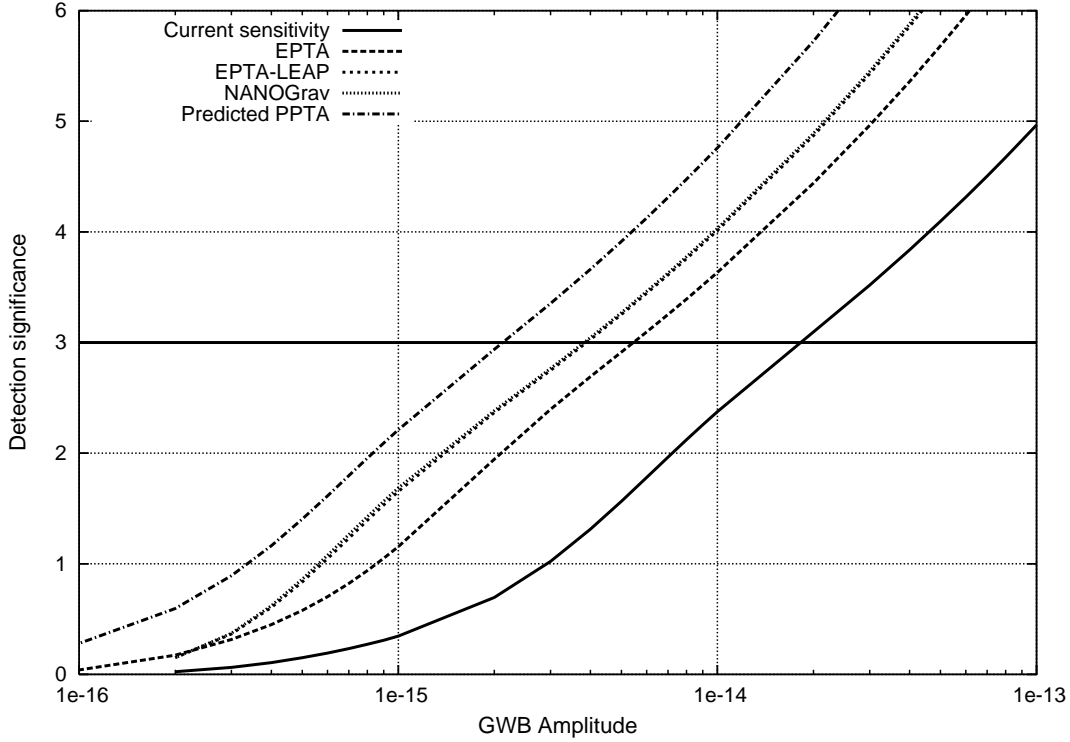}
  \caption{Sensitivity curves for different PTA efforts. Note the
    ``NANOGrav'' and ``EPTA -- LEAP'' curves are practically
    coincident. Gravitational waves are predicted to exist in the
    range $10^{-15}-10^{-14}$. See text and Table \ref{tab:PTAs} for
    more information.}
  \label{fig:PTAs}
\end{figure*}

The sensitivity curves presented in Figure \ref{fig:PTAs} seem to
justify cautious optimism for GWB detection through PTA experiments on
timescales of five to ten years, provided current models of GWBs are
correct. While none of the curves in Figure \ref{fig:PTAs} reach the
minimum predicted GWB Amplitude of $10^{-15}$ at a
detection-significance level of three, their sensitivity can be
expected to increase up to an order of magnitude through extension of
the campaigns to a decade-long timescale, as illustrated by the
difference between the ``Predicted PPTA'' and ``Current sensitivity''
curves. The GWB predictions may, however, change if other effects such
as black-hole binary stalling occur. The models do, furthermore, rely
on a substantial number of poorly determined input parameters, such as
what fraction of galaxy growth happens by merging \citep{svv09}. Since
only the merging of galaxies results in binary black holes and hence
contributes to the GWB, this mass fraction is crucial for any reliable
prediction of GWB strength.

As explained in \S\ref{sec:stab:disc}, the temporal systematic
contribution to the rms, $\sigma_{\tau}$, is a conservative upper
limit to the ultimate residual rms. In this analysis of PTA efforts,
however, we have used the value of $80\,$ns as a hard lower limit
on the timing rms, $\sigma$. This limits the potential for reduction
of the rms and explains the equivalence of the NANOGrav and EPTA--LEAP
efforts. Finally, the strong dependence on the timescale, $T$, of the
project underscores the importance of timing stability analysis over
much longer time spans and continued observing. While our $\sigma_{\rm
  z}$ analysis on PSR J1713+0747 provides the first evidence for high
timing stability over timescales beyond ten years, such timing
stability must still be demonstrated for many more MSPs.

\subsection{Future PTA Projects}\label{sec:FutPTAs}
With the completion of the Square Kilometre Array (SKA) pathfinders
expected in three years time, we consider the potential of the
Australian SKA Pathfinder (ASKAP), the South African Karoo Array
Telescope (MeerKAT) and the Chinese Five hundred meter Aperture
Spherical Telescope (FAST) for PTA programmes. ASKAP is primarily
designed for H$\,${\sc i} surveys and therefore sacrifices point
source sensitivity for a wide field of view, whereas MeerKAT's design
is better suited for point source sensitivity over a more limited
field of view. FAST is an Arecibo-type single dish with a total
diameter of 500\,m of which 300\,m is illuminated, resulting in a
substantially larger sky coverage than is possible with Arecibo. The
expected architecture for these telescopes is listed in Table
\ref{tab:PTAs} - notice we assume phase-coherent combination of the
signals of all ASKAP and MeerKAT dishes, effectively resulting in a
single telescope of diameter 107\,m for MeerKAT and 76\,m for ASKAP.

\begin{figure*}
  \includegraphics[width=14cm]{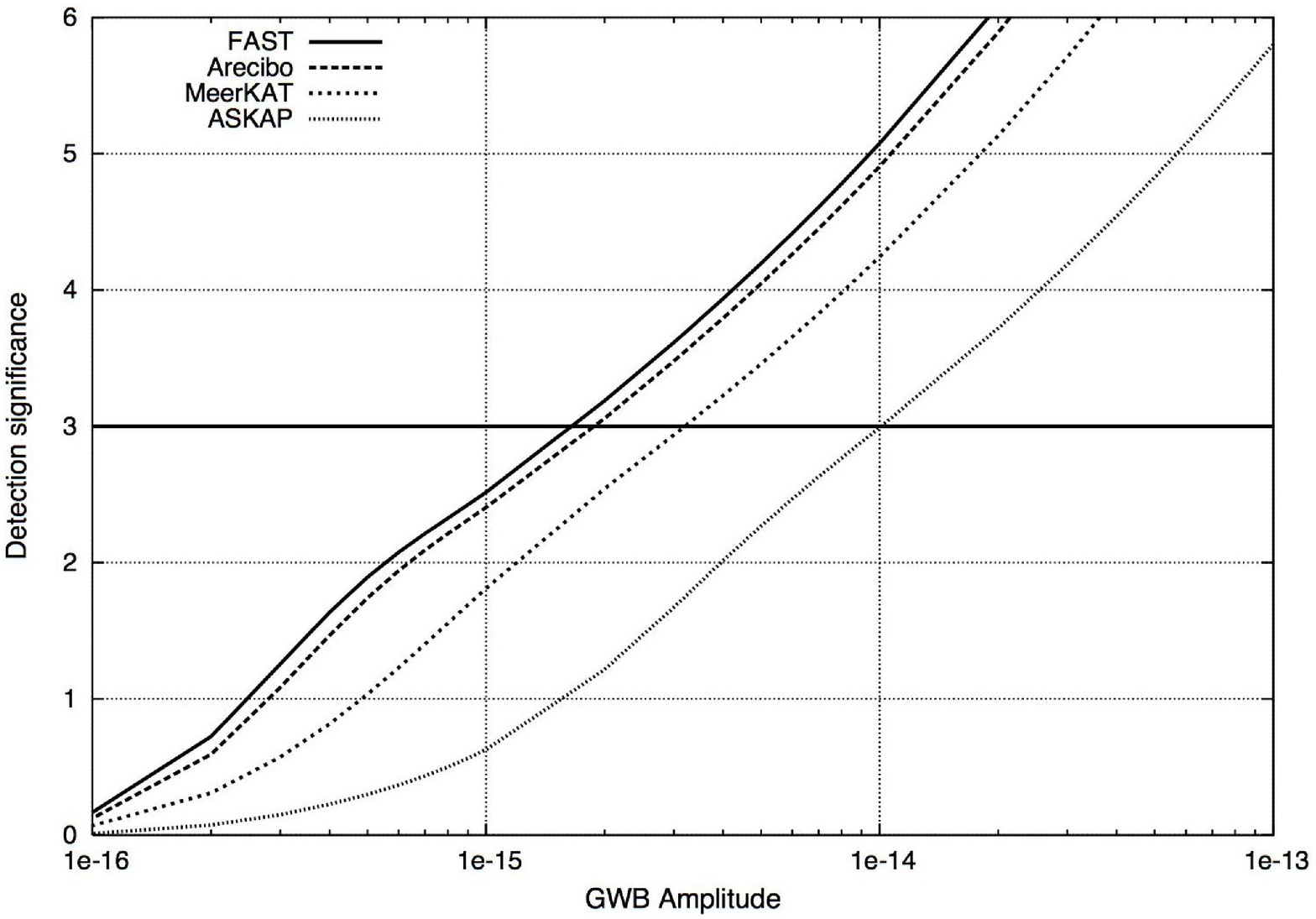}
  \caption{Sensitivity curves for the two main SKA pathfinders,
    Arecibo and FAST. Gravitational waves are predicted to exist in
    the range $10^{-15}-10^{-14}$. See discussion in \S\ref{sec:PTA}
    and Table \ref{tab:PTAs} for more information.}
  \label{fig:FutPTAs}
\end{figure*}

The resulting sensitivity curves are drawn in Figure
\ref{fig:FutPTAs}, along with a hypothetical curve for the most
sensitive telescope currently operational, the Arecibo radio
telescope. This figure clearly shows the advantage MeerKAT holds over
ASKAP for PTA work, in number of dishes, bandwidth and system
temperature. The sensitivity of Arecibo is much higher than that of
either interferometric prototype and is just inferior to FAST. As for
the NANOGrav and EPTA-LEAP projects analysed earlier, the advantage of
FAST over Arecibo is strongly limited by the bound of 80\,ns we
imposed on the achievable rms residual.

Note the usefulness of Arecibo for PTA work is limited by the
restricted sky coverage and hence available pulsars. While both
MeerKAT and ASKAP can see large parts of the Southern sky, the sky
coverage of Arecibo as well as the short transit time make an
exclusively Arecibo-based PTA practically impossible; however, its
potential as part of a combined effort (Figure \ref{fig:PTAs}) or in a
global PTA, is undeniable if the level of systematic errors is small
compared to the radiometer noise. As for any Northern telescope, the
usefulness of FAST will mostly depend on the discovery of good timing
MSPs at positive declinations.

\begin{table*}
  \begin{center}
	\begin{minipage}{17.5cm}
	  \caption{Assumed parameters for future and ongoing PTA
      efforts. Besides the names of the different PTAs, the columns
      contain the number of telescopes $N_{\rm Tel}$, the observing
      bandwidth $B$, the telescope diameter $D$, aperture efficiency
      $\eta$, system temperature $T_{\rm sys}$, observing regularity
      and the duration of the project, $T$.}
	  \label{tab:PTAs}
	  \begin{tabular}{lccrcclr}
		\hline 
        PTA & \multicolumn{1}{c}{$N_{\rm Tel}$} &
		\multicolumn{1}{c}{$B$} & \multicolumn{1}{c}{$D$} & $\eta$ &
    $T_{\rm sys}$ & Observing
		& \multicolumn{1}{c}{$T$} \\
        name & \multicolumn{1}{l}{} & \multicolumn{1}{c}{(MHz)} &
		\multicolumn{1}{c}{(m)}& & (K) & regularity &
		\multicolumn{1}{c}{(yrs)} \\
        \hline
        Current  & 1 & 64 & 64 & 0.6 & 25 & weekly & 5 \\ 
        Predicted PPTA & 1 & 256 & 64 & 0.6 & 25 & weekly & 10 \\
        NANOGrav & 2 & 256 & 305; 100 & 0.5; 0.7 & 30; 20 & monthly &
        5 \\
        EPTA     & 5 & 128 & 100 & 0.7 & 30 & monthly & 5\\
        EPTA - LEAP &
		1\footnote{Under the LEAP initiative, five 100\,m-class
		telescopes will be combined into an effective 224\,m single
		telescope.} & 128 & 224 & 0.7 & 30 & monthly & 5 \\
        \\ 
        Arecibo & 1 & 512 & 305 & 0.5 & 30 & two-weekly & 5 \\
        FAST\footnote{\citet{nan06,jng08}} & 1 & 400 & 500 & 0.36 & 20
        & two-weekly & 5 \\
        ASKAP\footnote{http://www.atnf.csiro.au/projects/askap/specs.html}
        & 40& 256 & 12 & 0.8 & 50 & weekly & 5 \\
        MeerKAT\footnote{\citet{jon07}
        } & 80 & 512 & 12 & 0.7 & 30 & weekly & 5\\
        \hline
	  \end{tabular}
	\end{minipage}
  \end{center}
\end{table*}

\section{Conclusions}\label{sec:conclusions}
We have presented the first long-term timing results for the 20 MSPs
constituting the Parkes Pulsar Timing Array (PPTA). We have shown that
only PSR J1939+2134 has timing instabilities that limit its use for
long-term GWB efforts, while the PSR J1713+0747 data already
demonstrate the requirements for GWB detection on a timescale of ten
to fifteen years are achievable. Overall, the timing stability of the
investigated MSPs was found to be encouraging even though potential
timing instabilities were detected in four pulsars (in addition to PSR
J1939+2134).

It was demonstrated that even on our most precisely timed MSPs, white
noise is a dominant contribution, suggesting that our residual rms
will be much improved with current wide-bandwidth systems. We placed a
conservative upper limit of $\sim$80\,ns on intrinsic timing
instabilities that will ultimately limit the residual rms. We
interpreted this result in the context of ongoing and future pulsar
timing array projects, demonstrating the realistic potential for GWB
detection through pulsar timing within five years to a decade,
provided technical and data reduction developments evolve as
expected. For PTA efforts in the Northern hemisphere, the discovery of
bright and stable MSPs in the Northern sky will be crucial. Given the
location of currently known MSPs, the prospects of the MeerKAT SKA
pathfinder as a gravitational-wave detector are found to be
particularly good.

\appendix
\section{PTA Sensitivity}\label{app:Sens}
In this Appendix we derive a simplified formalism for estimating the
sensitivity of a pulsar timing array (PTA) to a stochastic and
isotropic gravitational-wave background (GWB) of given amplitude,
$A$. This derivation produces results equivalent to those resulting
from equation (14) of \citet{jhlm05}, but is more readily implemented
and inherently treats optimal weighting (or prewhitening) of the
pulsar power spectra.

The detection statistic is the sample cross-covariance of the
residuals of two pulsars $i$ and $j$, separated by an angle
$\theta_{i,j}$:
\begin{equation}\label{eq:R}
  R(\theta_{i,j}) = \frac{1}{N_{\rm s}} \sum_{t = 0}^{T}
  r_{i}(t)\times r_{j}(t)
\end{equation}
(where $r_{i}(t)$ is the residual of pulsar $i$ at time $t$, $N_{\rm
  s}$ is the number of samples in the cross covariance and $T$ is the
data span). The expected value of $R(\theta_{i,j})$ is the covariance
of the clock error, which is 100\% correlated, plus the cross
covariance of the GWB, $\sigma_{\rm GW}^2 \zeta(\theta_{i,j})$. The
clock error can be included in the fit, but one must also include its
variance in the variance of the detection statistic. It is better to
estimate the clock error and remove it, which also removes its ``self
noise''. So in subsequent analysis we neglect clock noise and effects
of possible Solar System ephemeris errors. We
model the pulsar timing residuals as a GWB term and a noise term:
$r(t) = r_{\rm GW}(t) + r_{\rm N}(t)$, with variances
$\sigma_{\rm G}^2$ and $\sigma_{\rm N}^2$. $\zeta(\theta_{i,j})$ is
the cross-correlation curve predicted by \citet{hd83}, as a function
of the angle between the pulsars, $\theta_{i,j}$:
\[
\zeta(\theta_{i,j}) = \frac{3}{2}x \log x - \frac{x}{4} + \frac{1}{2}
\]
in which $x=(1-\cos{\theta_{i,j}})/2$.

Since the detection significance will be limited by the variance in
the sample cross covariance, we consider 
\begin{multline}
  {\rm Var}(R(\theta_{i,j})) \\
  \shoveleft{= {\rm Var}\left(\sum\left( 
	(r_{{\rm GW},i}+r_{{\rm N},i})(r_{{\rm GW},j}+r_{{\rm
  N},j})/N_{\rm s}\right)\right)}\\	
  \shoveleft{= \sigma_{{\rm G},i}^2 \sigma_{{\rm G},j}^2 
	            \frac{(1+\zeta(\theta_{i,j})^2)}
				     {N_{\rm s}}
    + \frac{\sigma_{{\rm N},i}^2 \sigma_{{\rm G},j}^2 +
	        \sigma_{{\rm G},i}^2 \sigma_{{\rm N},j}^2}
	       {N_{\rm s}}
    + \frac{\sigma_{{\rm N},i}^2 \sigma_{{\rm N},j}^2}{N_{\rm s}}}.\\
\end{multline}
After prewhitening this becomes (notice our notation $\sigma_{\rm PW}
= \varrho$):
\begin{multline}\label{eq:VarSum}
  {\rm Var}(R_{\rm PW}(\theta_{i,j})) \\
  \shoveleft{= \varrho_{\rm G}^4 \frac{(1+\zeta(\theta_{i,j})^2)}{N_{\rm s}}
	+ \varrho_{\rm G}^2 \frac{(\varrho_{{\rm N},i}^2+
	  \varrho_{{\rm N},j}^2)}{N_{\rm s}} + \frac{\varrho_{{\rm N},i}^2 
	  \varrho_{{\rm N},j}^2}{N_{\rm s}}}.\\
\end{multline}
in which we have used $\varrho_{{\rm G},i}^2 = \varrho_{{\rm
G},j}^2 = \varrho_{{\rm G}}^2$, which will be proved shortly.

We derive the gravitational-wave power from equations (\ref{eq:h_c}) and
(\ref{eq:GWBPower}), for a GWB with spectral index $\alpha = -2/3$:
\begin{equation}\label{eq:PGWB}
  P_{\rm GWB}(f) = K (f/f_{\rm ref})^{-13/3},
\end{equation}
with $K$ a constant proportional to the amplitude of the GWB and
$f_{\rm ref} = 1\,yr^{-1}$.

Defining the corner frequency, $f_{\rm c}$, as the frequency at which
the gravitational wave power equals the noise power, enables us to use
equation (\ref{eq:PGWB}) to determine the noise power: $P_{\rm Noise} = K
(f_{\rm c}/f_{\rm ref})^{-13/3}$. 

As illustrated by \citet{jhlm05}, the steep spectral index of
GWB-induced residuals implies that large gains in sensitivity can be
achieved through optimal prewhitening of the data. Assessment of the
variance of both the GWB and noise components of the residuals after
prewhitening, can most easily be done through integration of the
spectral powers, multiplied by the whitening filter, $W(f)$, which is
a type of Wiener filter, designed to minimize the error in the
estimation of $\sigma_{\rm G}$ and is of the form: 
$W(f) = P_{\rm GWB}/(P_{\rm GWB}+P_{\rm Noise})^2$. Rescaling the
weighting function thus defined, we get:
\begin{equation}\label{eq:Weights}
  W(f) = C \frac{\big(f/f_{\rm ref}\big)^{-13/3}}
  {\big(1+(f/f_{\rm c})^{-13/3}\big)^2}
\end{equation}
with $C$ a normalisation constant chosen for convenience to be:
\begin{equation}\label{eq:C}
  C = \Bigg(\sum_f{\frac{\big(f/f_{\rm ref}\big)^{-26/3}}{\big(1+(f/f_{\rm
  c})^{-13/3}\big)^2}}\Bigg)^{-1}
\end{equation}
The prewhitened variances then become:
\begin{eqnarray}
  \varrho_{\rm G}^2 &=& \sum_f K (f/f_{\rm ref})^{-13/3} C \frac{
  (f/f_{\rm ref})^{-13/3}}{\big(1+(f/f_{\rm c})^{-13/3}\big)^2}\notag\\
  &=& K \label{eq:sigmaG}\\
  \varrho_{\rm N}^2 &=& \sum_f K (f_{\rm c}/f_{\rm ref})^{-13/3} C \frac{
	(f/f_{\rm ref})^{-13/3}}{\big(1+ (f/f_{\rm c})^{-13/3}\big)^2}\notag\\
  &=& K C \sum_f{\frac{\big(f_{\rm c} f/f_{\rm
  ref}^2\big)^{-13/3}}{\big(1+(f/f_{\rm c})^{-13/3}\big)}}
\end{eqnarray}
which justifies our choice for $C$ and shows that, based on our
weighting scheme, $\varrho_{{\rm G},i}^2 = \varrho_{{\rm G},j}^2 = K$, as
used earlier.

Since the spectra are effectively bandlimited to $f_{\rm c}$ after
prewhitening, both the GWB and noise will have the same number of
degrees of freedom, namely: $N_{\rm dof} = 2 T_{\rm obs}f_{\rm c}-1$,
where $T_{\rm obs}$ is the length of the data span and therefore the
inverse of the lowest frequency, implying there are $T_{\rm obs}f_{\rm
  c}$ independent frequencies measured below $f_{\rm c}$. Since each
frequency adds a real and imaginary part, there are twice as many
degrees of freedom as there are independent frequency samples;
quadratic fitting removes a single degree of freedom from the
total. Notice that $\sqrt{N_{{\rm dof},i}N_{{\rm dof},j}}$ is the
number of independent samples in the cross-covariance spectrum and
therefore replaces $N_{\rm s}$ in equations (\ref{eq:R}) and
(\ref{eq:VarSum}).

The optimal least-squares estimator for $K$ (and hence for the
amplitude of the GWB), based on a given set $R_{\rm PW}(\theta_{i,j})$
with unequal errors, is (from equations (\ref{eq:R}) and
(\ref{eq:sigmaG})) :
\begin{equation}\label{eq:Estimator}
  \tilde{K} = \frac{\sum{R_{\rm PW}(\theta_{i,j})\zeta(\theta_{i,j})}
	/{\rm Var}(R_{{\rm
	PW,}i,j})}{\sum{\zeta(\theta_{i,j})^2/{\rm Var}(R_{{\rm PW,}i,j})}}
\end{equation}
The variance of this estimator is:
\begin{equation}\label{eq:EstVar}
  {\rm Var}(\tilde{K}) = 
  \frac{1}{\sum{\zeta(\theta_{i,j})^2/{\rm Var}(R_{{\rm PW,}i,j})}}
\end{equation}

We can now write the expected signal-to-noise of a given timing array
as the square root of the sum over all pulsar pairs of equation
(\ref{eq:sigmaG}) divided by the square root of equation (\ref{eq:EstVar})
\begin{equation}
  S = \sqrt{\sum_{i=1}^{N_{\rm psr}-1} \sum_{j=i+1}^{N_{\rm psr}}
    \frac{\varrho_{\rm G}^4 \zeta^2 \sqrt{N_{{\rm dof},i}N_{{\rm
            dof},j}}}{\varrho_{\rm G}^4 (1+\zeta^2)+\varrho_{\rm
        G}^2(\varrho_{{\rm N},i}^2+\varrho_{{\rm
          N},j}^2)+\varrho_{{\rm N},i}^2 \varrho_{{\rm N},j}^2}}.
\end{equation}
Rewriting leads to:
\begin{equation}
  S = \sqrt{\sum_{i=1}^{N_{\rm psr}-1}\sum_{j=i+1}^{N_{\rm psr}}
	\frac{\zeta^2 \sqrt{N_{{\rm dof},i}N_{{\rm dof},j}}}
		 {1+\zeta^2+
		   \left(\varrho_i^{\prime}\right)^2+
		   \left(\varrho_j^{\prime}\right)^2+
		   \left(\varrho_i^{\prime}
		   \varrho_j^{\prime}\right)^2}}
\end{equation}
where $\varrho_i^{\prime} = \varrho_{{\rm N},i}/\varrho_{\rm G}$.
\vspace{0.5cm}

{\bf \noindent Acknowledgments.}

The Parkes Observatory is part of the Australia Telescope which is
funded by the Commonwealth of Australia for operation as a National
Facility managed by CSIRO. We thank the staff at Parkes Observatory
for technical assistance and dedicated help over many years. We also
acknowledge the support of the large number of collaborators and
students who have assisted in the acquisition of the data presented in
this paper over the last 14 years. JPWV acknowledges financial support
provided by the Astronomical Society of Australia (ASA). XPY is
supported by the National Natural Science Foundation (NNSF) of China
(10803004) and by Natural Science Foundation Project CQ CSTC
(2008BB0265).

\bibliographystyle{mn2e}
\bibliography{journals,psrrefs,modrefs,crossrefs}

\end{document}